\documentclass[twocolumn,trackchanges,floatfix,twocolappendix,longbib]{aastex701}
\usepackage{booktabs}
\usepackage[english]{babel}
\usepackage{url}
\usepackage{natbib}
\usepackage{gensymb}
\usepackage{subfigure}
\DeclareUnicodeCharacter{2212}{-}
\usepackage{amsmath,bm}
\usepackage{longtable}
\usepackage{multirow}
\usepackage{threeparttable}
\usepackage{times}
\usepackage{enumitem}
\usepackage{lipsum}[1-2]
\usepackage{array}
\usepackage{tabularx}
\usepackage{float}

\newcommand{\rgc}{$r_{\rm GC}$}
\newcommand{\Rgc}{$R_{\rm GC}$}

\shorttitle{Estimating Fe and $\alpha$ Abundances of the M31 Halo}
\shortauthors{Hong et al.}
\begin{document}

\title{The Star Formation Histories of Different Components of M31's Halo \\ from Photometric Estimates of Fe and $\alpha$ Abundances}

\author[0000-0002-2453-0853]{Jihye Hong}
\affiliation{Department of Physics and Astronomy, University of Notre Dame, Notre Dame, IN 46556, USA}
\affiliation{Joint Institute for Nuclear Astrophysics -- Center for the Evolution of the Elements (JINA-CEE), USA}
\email{jhong5@nd.edu}

\author[0000-0001-6196-5162]{Evan N. Kirby}
\affiliation{Department of Physics and Astronomy, University of Notre Dame, Notre Dame, IN 46556, USA}
\email{ekirby@nd.edu}

\author[0000-0003-0398-639X]{Roman Gerasimov}
\affiliation{Department of Physics and Astronomy, University of Notre Dame, Notre Dame, IN 46556, USA}
\email{rgerasim@nd.edu}

\author[0000-0002-9053-860X]{Masashi Chiba}
\affiliation{Astronomical Institute, Tohoku University, 6-3 Aoba, Aramaki, Aoba-ku, Sendai, Miyagi 980-8578, Japan}
\email{chiba@astr.tohoku.ac.jp}

\author[0000-0002-3852-6329]{Yutaka Komiyama}
\affiliation{Department of Advanced Sciences, Faculty of Science and Engineering, Hosei University, 3-7-2 Kajino-cho, Koganei, Tokyo 184-8584, Japan}
\email{komiyama@hosei.ac.jp}

\author[0000-0001-8239-4549]{Itsuki Ogami}
\affiliation{Department of Advanced Sciences, Faculty of Science and Engineering, Hosei University, 3-7-2 Kajino-cho, Koganei, Tokyo 184-8584, Japan}
\affiliation{National Astronomical Observatory of Japan, 2-21-1 Osawa, Mitaka, Tokyo 181-8588, Japan}
\affiliation{The Institute of Statistical Mathematics, 10-3 Midoricho, Tachikawa, Tokyo, 190-8562, Japan}
\affiliation{Department of Physics, Graduate School of Science, Tokyo Metropolitan University, 1-1 Minami-Osawa, Hachioji-shi, Tokyo 192-0397, Japan}
\email{ogami@tmu.ac.jp}

\author[0000-0002-8079-6867]{Tiffany M. Tang}
\affiliation{Department of Applied and Computational Mathematics and Statistics, University of Notre Dame, Notre Dame, IN 46556, USA}
\email{ttang4@nd.edu}

\author[0000-0003-4573-6233]{Timothy C. Beers}
\affiliation{Department of Physics and Astronomy, University of Notre Dame, Notre Dame, IN 46556, USA}
\affiliation{Joint Institute for Nuclear Astrophysics -- Center for the Evolution of the Elements (JINA-CEE), USA}
\email{Timothy.C.Beers.5@nd.edu}

\author[0000-0001-5297-4518]{Young Sun Lee}
\affiliation{Department of Astronomy and Space Science, Chungnam National University, Daejeon 34134, Republic of Korea}
\email{youngsun@cnu.ac.kr}

\submitjournal{ApJ}

\begin{abstract}
We present the chemical-abundance distributions and star-formation histories of the M31 halo components using photometric Fe and $\alpha$ abundance estimates derived from the PAndAS broad-band $g$ and $i$ filters and the Subaru/HSC $NB515$ narrow-band filter. We selected about 76,400 candidate M31 giant stars and constructed a synthetic photometric training dataset based on model isochrones. We estimated elemental abundances and errors using random forest regression models. We identify various halo components, including about 32,500 stars in the ``smooth" 
halo (where ``smooth" refers to 
the complete sample with all the known substructures removed), and infer their star-formation histories from comparisons of their abundance distributions. From the radial abundance gradients, we find that [Fe/H] decreases while [$\alpha$/Fe] increases toward the outer halo. Including substructures further enhances the radial abundance contrast, suggesting that the smooth halo stars are accreted at relatively early times, before the progenitor galaxies produce large numbers of Type~Ia supernovae. Both M31 and the Milky Way show similar radial gradient trends, but M31 exhibits higher [Fe/H] and lower [$\alpha$/Fe] values overall, suggesting that the smooth 
halo of M31 may possess a larger fraction of relatively younger stars associated with more recent merger events. Future photometric observations with additional narrow-band filters and larger spectroscopic samples are expected to provide a better understanding of the formation and evolution history of M31. 

\end{abstract}


\section{Introduction}\label{sec:introduction}
The Andromeda galaxy, M31, located at a distance of 783 kpc \citep{McConnachie2005}, offers a global view for investigating the evolutionary history of a large spiral galaxy in the Local Group. Our external vantage point of M31, in some ways, makes it easier to study its stellar halo than that of the Milky Way. The star formation of M31's components (e.g., disk and halo) has been measured in increasing detail over the past two decades.  Studies of star formation in the disk include \citet{Williams2003} and \citet{Kang2009}, as well as the Panchromatic Hubble Andromeda Treasury (PHAT) survey \citep{Dalcanton2012}. More recent spectroscopic studies include \citet{Escala2020a}, which reported the [Fe/H] vs.\ [$\alpha$/Fe] distribution (the ``Tinsley--Wallerstein diagram'') in the disk at a projected distance of \Rgc $\sim$ 26 kpc from the center of M31. By observing planetary nebulae, \citet{Bhattacharya2019, Bhattacharya2022} found that the thin and thick disks of M31 extend to galactocentric radii of \rgc$\sim 14$ and 30 kpc, respectively, and \citet{Kobayashi2023} reported the [$\alpha$/Fe] vs.\ [Fe/H] distribution in this region, along with its bimodality in the context of galactic chemical evolution models.

In this work, we focus on the stellar halo, which contains many old stars that sample the early formation and evolution history of M31 \citep{McConnachie2018}. Spectroscopic analyses carried out so far include, for example, \citet{Gilbert2012}, who aimed to obtain chemical and dynamical measurements of M31’s dwarf satellites through the Spectroscopic and Photometric Landscape of Andromeda’s Stellar Halo (SPLASH) survey. Using medium-resolution ($R=\lambda$/$\Delta\lambda \sim 6,000$) Keck II/DEIMOS observations, they measured over 1,500 red giant branch (RGB) halo stars out to \Rgc $\sim$ 175 kpc and reported that the surface brightness profile \citep{Kalirai2006, Koch2008, Tanaka2010} follows a power law with an index of $-2.2$. 
In addition, they suggested that although most stellar halo formation simulations predict a sharp decline in surface brightness beyond about 100 kpc, only one simulated halo in \citet{Bullock2005}, which experienced numerous recent low-mass accretion and disruption events, did not show this behavior, implying that M31 may have undergone a similar accretion history.

Studies that have reported [Fe/H] and [$\alpha$/Fe] measurements include \citet{Vargas2014}, who analyzed four RGB stars in the outer halo region at \Rgc\ = 70 to 140 kpc from the \citet{Gilbert2012} sample, using higher signal-to-noise (S/N) spectra obtained from follow-up DEIMOS observations. They found [Fe/H] ranging from $-1.4$ to $-2.2$ and a mean of ${\rm [\alpha/Fe]} = +0.20 \pm 0.20$ dex. Furthermore, including the two stars overlapping with \citet{Vargas2014}, \citet{Gilbert2020} reported 23 [Fe/H] and nine [$\alpha$/Fe] measurements using spectral synthesis at \Rgc\ = 43 to 165 kpc. The median [Fe/H] values spanned about $-1.3$ to $-2.1$, showing a negative [Fe/H] gradient. They reported a mean of ${\rm [\alpha/Fe]} = +0.30 \pm 0.16$ dex.

In addition, several studies have reported abundances for halo stars at smaller projected radii. For example, \citet{Escala2019} applied spectral synthesis to low-resolution ($R\sim$2,500 at 7,000 \AA) DEIMOS spectroscopy for the region at \Rgc\ = 23 kpc along the minor axis of M31. They considered 11 RGB stars in this region as part of the smooth inner halo -- where ``smooth" refers to a kinematically hot ($\sigma_v = 132$ km s$^{-1}$) spheroid component with no detectable substructures -- and obtained a mean [Fe/H] = $-1.59 \pm 0.56$ dex and a mean [$\alpha$/Fe] = $+0.49 \pm 0.29$ dex, consistent with \citet{Gilbert2007}. They suggested that the low metallicity and highly enhanced [$\alpha$/Fe] likely reflect the early accretion and disruption of dwarf galaxies with high star-formation rates and truncated star-formation histories (SFHs) during the formation of M31’s halo. Subsequently, \citet{Escala2020a} reanalyzed the sample at \Rgc\ = 23 kpc using a methodology consistent with those adopted for the other fields, along with a additional radial velocity membership cut, and reported  a mean $\rm [Fe/H] = -1.54 \pm 0.14$ dex and a mean $\rm [\alpha/Fe]= +0.43^{+0.11}_{-0.12}$ dex. Furthermore, \citet{Escala2020b} analyzed 128 RGB stars with DEIMOS spectra over \Rgc\ = 8 to 34 kpc and reported a negative [Fe/H] gradient of $-0.025\pm0.002$ dex~kpc$^{-1}$, a mean $\rm [Fe/H] = -1.08\pm0.04$ dex, and a mean $\rm [\alpha/Fe] = +0.40\pm0.03$ dex.

Most recently, \citet{Wojno2023} co-added low-S/N SPLASH spectra following the methods of \citet{Yang2013} and \citet{Wojno2020}, and analyzed the smooth halo at \Rgc\ = 8 to 177 kpc, spanning from the inner halo studied by \citet{Escala2020b} to the outer halo studied by \citet{Vargas2014} and \citet{Gilbert2020}. Using a total of 155 individual and co-added measurements, they obtained an [Fe/H] gradient of $-0.0075\pm0.0012$ dex~kpc$^{-1}$ and an [$\alpha$/Fe] gradient of $+0.0030\pm0.0009$ dex~kpc$^{-1}$; the mean [Fe/H] decreased from $-1.27\pm0.21$ dex at \Rgc\ $<50$ kpc to $-1.94\pm0.20$ dex at \Rgc\ $>50$ kpc, whereas the mean [$\alpha$/Fe] increased from $+0.29\pm0.08$ dex to $+0.43\pm0.12$ dex.

There are also many studies of the M31 halo using photometry.
For instance, using SPLASH survey data, \citet{Gilbert2014} reported photometric [Fe/H], ranging from about $-0.4$ to $-2.0$ (slightly higher than spectroscopic measurements by \citet{Gilbert2020}), and a negative gradient toward the outer halo, with a slope of about $-0.0110\pm0.0007$ dex~kpc$^{-1}$ over \Rgc\ = 10 to 90 kpc. After excluding kinematically cold ($\sigma_v = 5$–50 km s$^{-1}$) substructures, the overall median [Fe/H] values appeared to be slightly lower, and the negative gradient still remained with a slope of $-0.0105\pm0.0013$ dex~kpc$^{-1}$. 

Additionally, \citet{McConnachie2009}, \citet{Ibata2014}, and \citet{McConnachie2018} analyzed the RGB population of M31’s halo at about 27 kpc $<$ \Rgc\ $<$ 150 kpc using the Pan-Andromeda Archaeological Survey (PAndAS) photometry with $g$ and $i$ broad-band filters. Using 13 Gyr Dartmouth isochrones \citep{Dotter2008} to estimate photometric [Fe/H], \citet{Ibata2014} divided stars with 0.8 $< (g − i)_0 <$ 1.8 and $i_0 <$ 23.5 mag into four metallicity bins: ($−2.5$, $−1.7$), ($−1.7$, $−1.1$), ($−1.1$, $−0.6$), and ($−0.6$, $0.0$). For each metallicity bin, they examined the spatial density distribution, sphericity, and power-law index both for all stars and for the smooth halo, excluding streams and substructures (see their Figure 10). They found that the halo stars are distributed almost spherically and that the absolute value of the power-law index increases with increasing metallicity, although this increase is smaller for the smooth halo. Furthermore, by examining the [Fe/H] distributions of all stars across five \Rgc\ intervals, they showed that the mean [Fe/H] decreases from roughly $-0.7$ in the inner halo to $-1.5$ in the outer halo, while for the smooth inner-halo sample, the mean [Fe/H] ranges from about $-0.9$ to $-1.6$.

The most recent analyses of photometric surveys used the Hyper Suprime-Cam \citep[HSC;][]{Miyazaki2012, Miyazaki2018, Furusawa2018, Kawanomoto2018, Komiyama2018b} on the Subaru Telescope. For example, \citet{Komiyama2018a} analyzed observations of the northwestern I \citep[NW I;][]{McConnachie2009, Richardson2011} stream of M31, covering an area of 9.2 deg$^2$, and calculated distances to red clump (RC) stars. They reported that the diffuse southern part of this stream extends toward us by about 20 kpc compared to the more distant northern region. In this study, the RC and RGB stars with $g \sim 26.3$ mag were selected using HSC broad-band $g$ and $i$ filters, along with the narrow-band $NB515$ filter, which targets surface gravity-sensitive MgH and Mg I~\textit{b} absorption features, similar to the use of the DDO51 filter \citep{Majewski2000}.

In addition, \citet{Ogami2025} utilized photometric data from the $g$ and $i$ bands of PAndAS and the $NB515$ narrow-band of HSC\@. They selected stars with $(g - i)_0 < 2.5$ and $i_0 < 23$ mag after applying Galactic extinction corrections, extending out to a \Rgc\ of 120 kpc. By using the $(g-i)_0$ vs. $(NB515-g)_0$ color-color diagram to derive RGB probabilities and combining them with the spatial distribution of foreground dwarf stars as a function of Galactic latitude \citep{Tanaka2010}, they selected RGB samples and employed 13 Gyr Dartmouth isochrone to analyze the distribution of [Fe/H] estimates. They determined a mean metallicity of [Fe/H]$ = -0.83\pm0.54$ dex and reported a metallicity gradient across four radial bins of 0 to 2, 2 to 4, 4 to 6, and 6 to 10$\degree$ (corresponding to \Rgc\ $\sim$ 24, 54, 81 kpc, and the outermost region, assuming a distance of 776 kpc to M31 \citep{Dalcanton2012}), with mean [Fe/H] values ranging from $-0.80$ to $-1.05$. They also confirmed a negative gradient along the minor axes from the center toward the southeast and northwest regions, respectively. Moreover, they examined the surface brightness profiles in each metallicity bin, in a manner similar to \citet{Ibata2014}, and suggested that these brightness profiles, along with the inhomogeneous metallicity distributions, reflect dynamically unmixed merger remnants; the inner halo is dominated by more massive metal-rich merger remnants, while the outer halo mainly consists of smaller metal-poor merger remnants, consistent with predictions of the hierarchical $\Lambda$CDM paradigm.

\begin{figure*}
    \centering
    \includegraphics[width=0.9\textwidth]{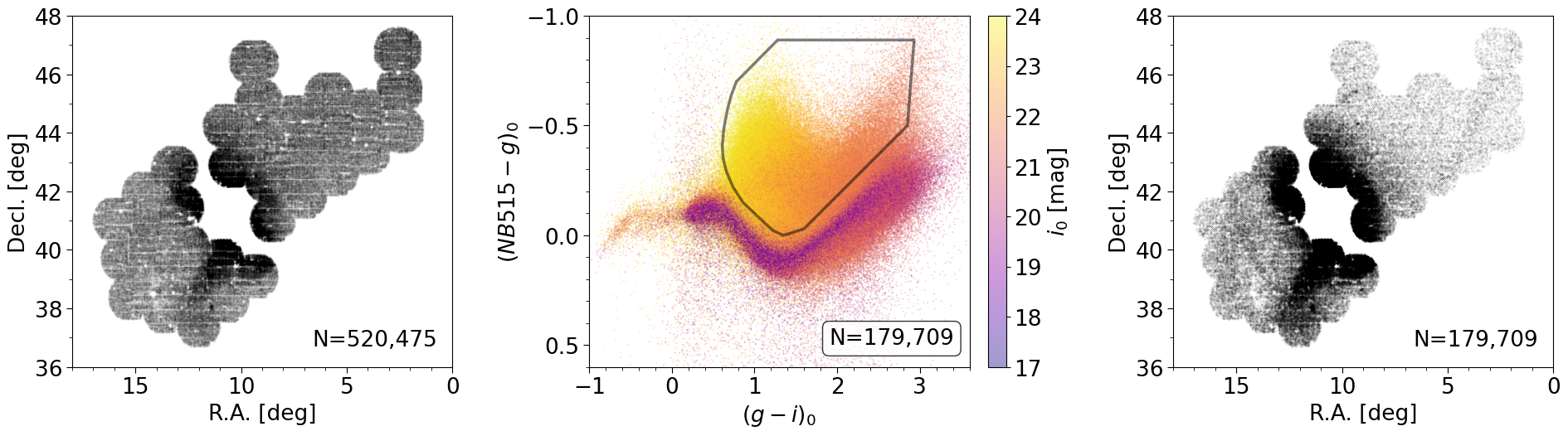}
    \caption{Left: The equatorial coordinates in R.A. and Decl. of the M31 halo sample. A cut on photometric errors of $\leq 0.10$ mag for the broad-band filters and $\leq 0.15$ mag for $NB515$ was applied. Middle: A color–color diagram, color-coded by $i_0$ magnitudes. The polygon marks the giant candidate stars with their number indicated in the lower right corner. Right: Same as the left panel, but for the giant candidates selected within the polygon.} 
    \label{Fig:ccd_select_sample}
\end{figure*}

In this study, we use the M31 data from \citet{Ogami2025} to investigate the overall chemical-abundance distribution of the M31 halo. Giant stars are selected based on $(g - i)_0$ and $(NB515 - g)_0$ colors, and the [Fe/H] and [$\alpha$/Fe] abundances of about 76,400 stars are estimated using a Random Forest (RF) regressor similar to \citet{Hong2025}. We then examine the global spatial chemical distribution and extract substructures such as dSphs and streams within the observation fields to compare their distributions with the smooth 
halo regions and infer their distinct SFHs. Finally, we compare these results with those of the Milky Way's halo.

\begin{figure}
    \centering
    \includegraphics[width=0.4\textwidth]{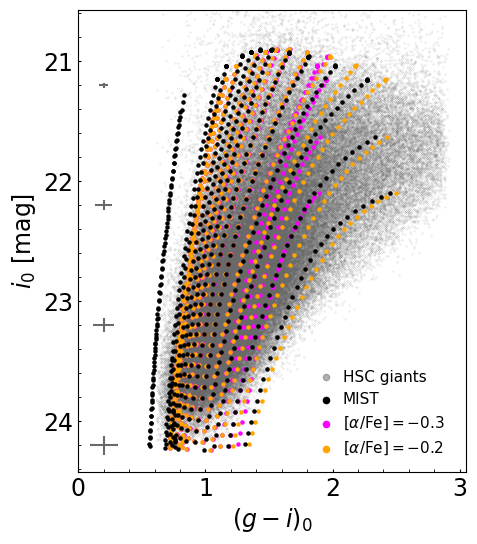}
    \caption{CMD of the M31 halo giant candidates selected by the polygon in gray dots. For stars at $(g - i)_0 = 1$ and in one-magnitude intervals of $i_0$, the mean $(g - i)_0$ errors, obtained in quadrature, and the mean $i_0$ errors are shown as error bars on the left. The 10 Gyr [$\alpha$/Fe] $=-0.2$ MIST isochrone is shown as black dots, and the derived $(g-i)_0$ colors for [$\alpha$/Fe] $=-0.3$ and $-0.2$, obtained by running Chemfit are shown as magenta and orange dots, respectively.}
    \label{Fig:cmd_mist_chemfit}
\end{figure}

\begin{figure}
    \centering
    \includegraphics[width=0.45\textwidth]{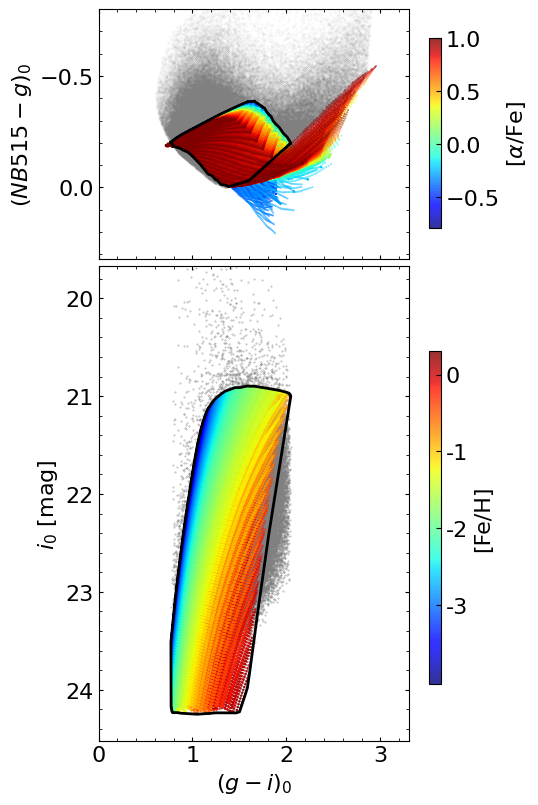}
    \caption{Top: A color-color diagram of the M31 halo giant candidates shown as gray dots, with the synthetic photometric data overplotted and color-coded by [$\alpha$/Fe]. The black line indicates the convex hull defining the selection of M31 halo giant sample and synthetic photometric data used in the bottom panel. Bottom: CMD of the M31 halo giant sample and the synthetic photometric data color-coded by [Fe/H], selected in the top panel. The black line shows the convex hull enclosing the synthetic photometric data in the CMD, and the M31 halo giants within this boundary were used for the abundance estimation.}
    \label{Fig:ccdcmdcut}
\end{figure}

The remainder of this paper is organized as follows. Section~\ref{sec:dataandmethods} describes the M31 halo photometric data used in our analysis, the synthetic photometric data for the training sample derived from isochrones, and the RF methods used for elemental abundance estimation. Section~\ref{sec:results} presents the model evaluation, error estimations, the color-color and color-magnitude diagram results, and the [Fe/H] vs.\ [$\alpha$/Fe] distribution. In Section~\ref{sec:discussion}, we show the overall spatial distributions of chemical abundances in M31 and compare the [Fe/H] vs.\ [$\alpha$/Fe] distributions of the substructures with those of the smooth 
halo, and compare the radial abundance gradients with previous studies of M31 as well as with the Milky Way. A summary is provided in Section~\ref{sec:summary}.

\section{Data and Methods}\label{sec:dataandmethods}
Here we describe the M31 halo photometric data adopted as the target sample for abundance estimation, the synthetic photometric sample constructed from isochrones, and the RF method used to estimate the Fe and $\alpha$ abundances.

\subsection{Photometric Sample}\label{sec:hscphot}
Similar to \citet{Ogami2025}, we used photometric data obtained from the PAndAS $g$ and $i$ broad-band filters and the Subaru/HSC $NB515$ narrow-band filter. PAndAS is a photometric survey designed to understand the structure and evolution of M31 and its satellite systems. It was conducted with MegaPrime/MegaCam on the 3.6 m Canada–France–Hawaii Telescope (CFHT), which has a one deg$^2$ field of view (FoV) and reaches depths of $g = 26.5$ mag (4,140–5,600 \AA) and $i = 25.5$ mag (7,020–8,530 \AA). Observations were carried out between 2008 and 2011, covering more than 400 degrees$^2$. The $NB515$ filter, centered at 5,150 \AA\ with a full width at half maximum (FWHM) of about 77 \AA\ \citep{Ogami2025}, reaches a depth of 25.7 mag and was used with the HSC on the 8.2 m Subaru Telescope. It has a FoV of about 1.8 degrees$^2$ and was used to observe about 50 degrees$^2$ of the M31 halo in 2015 and 2019 \citep{Komiyama2018b, Ogami2025}. The data reduction and Galactic extinction correction were performed by \citet{Ogami2025}. The final data set comprises about 1.52 million point-like sources classified as \texttt{extendedness = 0} in the catalog.

To improve data quality, we computed the inverse-variance-weighted mean of the $NB515$ magnitude errors for about 752,000 duplicate measurements based on R.A. and Decl., corresponding to roughly 330,000 stars. Then, we included only stars with broad-band and narrow-band errors below 0.10 and 0.15 mag, respectively. This resulted in a sample of about 520,500 stars. Their equatorial coordinates and the $(g - i)_0$ vs.\ $(NB515 - g)_0$ color–color diagram are shown in the left and middle panels of Figure~\ref{Fig:ccd_select_sample}, respectively.

The $NB515$ filter targets the Mg I~\textit{b} triplet at 5,180 \AA\ and the MgH band at 5,210 \AA\ \citep{Ogami2025}, which are sensitive to surface gravity, and is therefore not only useful for estimating Mg abundances but also effective at distinguishing giant stars with relatively low surface gravity from foreground Milky Way dwarfs by defining a polygon on the color–color diagram, as demonstrated in \citet{Komiyama2018b}, \citet{Okamoto2024}, \citet{Ogami2025}, and \citet{Hong2025}. We used a similar polygon-based selection to identify likely giant stars. 

Recently, \citet{Ding2025} estimated the probability that stars are M31 giants using a neural network and identified stars with probabilities greater than 0.9 as giant candidates. However, their selection includes more stars with lower $(NB515-g)_0$ values, extending into the disk-like population in the purple check-mark-shaped region at bright $i_0$ in the middle panel of Figure~\ref{Fig:ccd_select_sample}. Thus, we retained the polygon-based selection to minimize contamination from disk stars. Approximately 179,710 giant candidates within the polygon are shown in the middle panel, and their equatorial coordinates are shown in the right panel of Figure~\ref{Fig:ccd_select_sample}. Among the giant candidates selected by the polygon, 99.3\% were identified as giant candidates based on the probabilities of \citet{Ding2025}.

\subsection{Synthetic Photometric Sample}\label{sec:synphot}
Empirical spectroscopic observations of abundances for a large number of stars would be the ideal training dataset, as in \citet{Hong2025}. We found 23 DEIMOS spectroscopic data that have R.A. and Decl. positions overlapping within 1 arcsecond with the M31 halo giant candidates: 1 from \citet{Kirby2020} and 22 from \citet{Escala2020a}. However, 23 stars is far too few for accurate training. A larger complementary synthetic data set is required that provides magnitudes, colors, and abundances. Thus, we obtained 7, 10, and 13 Gyr MIST isochrones \citep{Choi2016} v2.5 in the CFHT system for $\rm [\alpha/Fe] = -0.2, 0.0, +0.2, +0.4$, and $+0.6$, which provide stellar parameters together with magnitudes in the $g$ and $i$ bands. We then applied cuts to the isochrones to match the $(g - i)_0$ and $M_{i_0}$ ranges of the M31 halo giants in the color-magnitude diagram (CMD), restricting the synthetic data to the RGB region. We adopted a distance modulus of 24.47 \citep{deVaucouleurs1991, McConnachie2005, McConnachie2012} to derive absolute magnitudes for the M31 halo sample.
Here, given the sensitivity of the $NB515$ filter to Mg abundance, we treated Mg as a representative $\alpha$ element and used the isochrone [$\alpha$/Fe] values in the subsequent analysis to estimate the [$\alpha$/Fe] abundance ratios of the M31 halo sample.

To estimate abundances, we required not only the $(g - i)_0$ and $M_{i_0}$ values from the isochrones but also the $(NB515 - g)_0$ color. This color was derived using Chemfit\footnote{https://github.com/Roman-UCSD/chemfit}, which is both a library of spectral models and software to determine best-fit stellar parameters by fitting observed spectra to those models. It was developed for the `{\= O}nohi`ula Subaru Prime-Focus Spectrograph (PFS), and its synthetic photometry method is described in \citet{Gerasimov2022}.
We first restricted the isochrone data to fall within the full parameter ranges available in the Chemfit ($T_{\rm eff}$ = 3,500 to 8,000 K, 
$\log g$ = 0.0 to 5.0, 
[Fe/H] = $-5.0$ to $+0.5$ and [$\alpha$/Fe] = $-2.0$ to +1.2) 
and then applied Chemfit to generate synthetic photometry that includes all colors.
When running Chemfit, we defined the [$\alpha$/Fe] values to span from $-0.8$ to $+1.0$ in steps of 0.1, assigning the available MIST isochrones to different [$\alpha$/Fe] ranges. Specifically, the [$\alpha$/Fe] = $-0.2$, 0.0, $+0.2$, $+0.4$, and $+0.6$ isochrones were adopted for synthetic [$\alpha$/Fe] ranges of $-0.8$ to $-0.2$, $-0.1$ to 0.0, $+0.1$ to $+0.2$, $+0.3$ to $+0.4$, and $+0.5$ to $+1.0$, respectively. In addition, the $g$ and $i$ filter transmission curves were taken from the CFHT/MegaCam filters provided by Chemfit, and that of $NB515$ was obtained from the Subaru Telescope HSC website\footnote{https://subarutelescope.org/Instruments/HSC/sensitivity.html}. All magnitudes were given in the AB system. We also assumed a radial velocity of $-300$ km s$^{-1}$ and adopted [C/Fe] = 0.0, with the zero point offset according to each star's log g and [Fe/H] following \citet{Placco2014}.

In Figure~\ref{Fig:cmd_mist_chemfit}, the CMD of the M31 halo giant sample is shown as gray dots and the 10 Gyr [$\alpha$/Fe] $=-0.2$ MIST isochrone CMD as black dots. Using this isochrone, the derived $(g-i)_0$ colors obtained from Chemfit with the isochrone $i_0$ are shown as magenta and orange dots for [$\alpha$/Fe] $=-0.3$ and $-0.2$, respectively. 
For the synthetic photometric data, [$\alpha$/Fe] $=-0.3$ appears to be somewhat limited in the bluest region compared with the MIST isochrones; indeed, we found that for $-0.8 \leq$ [$\alpha$/Fe] $\leq -0.3$, the Chemfit-derived colors are restricted to a less red range. 
This discrepancy at redder $(g-i)_0$ colors likely arises from Chemfit's use of ATLAS atmosphere models \citep{Castelli2003}, which are known to lose accuracy at low effective temperatures, $T_{\rm err} < 4,000$ K.

For the interpolation, we used DFInterpolator from the Python \texttt{isochrones.interp} module, with [$\alpha$/Fe], [Fe/H]$_{\rm init}$, which represents the initial [Fe/H] when the stars formed, and the equivalent evolutionary point (EEP) as the grid dimensions. We interpolated the synthetic photometric data by a factor of three and obtained interpolated values for the magnitudes and colors, constructing the final interpolated synthetic photometric dataset to be used as the training set for abundance estimation. The color-color diagram of this dataset, color-coded by [$\alpha$/Fe] and overplotted on the M31 halo giants shown as gray dots, is presented in the upper panel of Figure~\ref{Fig:ccdcmdcut}. We then used ConvexHull from the Python \texttt{scipy.spatial} module to select the region where the synthetic sequences remain well ordered in [$\alpha$/Fe]. The lower panel shows the CMD of the selected synthetic data and the M31 halo giant sample within this region, color-coded by [Fe/H]. We applied another ConvexHull in the $(g-i)_0$ and $i_0$ space to the synthetic data selected in the color-color space and retained only giants within this second convex hull. Therefore, the interpolated synthetic photometric training sets contain 233,943, 238,224, and 236,537 stars for 7, 10, and 13 Gyr, respectively, and the corresponding M31 halo giant samples contain 95,988, 96,737, and 97,481 stars. 
Below, we describe the 10 Gyr case as an example. 
The selection cuts based on the isochrone ranges used to define the final M31 giant sample here may depend somewhat on the adopted isochrone models.

\subsection{Methods}\label{sec:methods}
To estimate the abundances of [Fe/H] and [$\alpha$/Fe] of the M31 halo giant sample, we used a RF regressor \citep{Breiman2001} following the approach of \citet{Hong2025}. An RF is an ensemble machine learning model composed of multiple decision trees, each of which recursively performs binary splits to maximize the reduction in variance at each threshold. Each tree is trained independently using bootstrap-resampled samples randomly drawn from the training set. We adopted RF because it is known to be particularly well-suited for capturing complex and nonlinear relationships in the data compared to conventional statistical methods \citep{Breiman2001, Hastie2009, Biau2015}.

Considering that the distribution of stars in the CMD traces stellar populations from bluer, metal-poor stars to redder, metal-rich stars at a given age, we used the $M_{i_0}$, $(g-i)_0$, and $(NB515-g)_0$ values of the synthetic photometric data as input features to train an RF regression model to simultaneously estimate [Fe/H] and [$\alpha$/Fe] for the M31 halo giant sample.
Eighty percent of the synthetic photometric data were used for training, and the remaining twenty percent were reserved as a validation (test) set. 
For hyperparameter optimization, due to the large size of the training sample, we randomly selected 1,000 stars from the training data set and used five-fold cross-validation. The optimized hyperparameters were the number of trees, \texttt{n\_estimators} = 200, and the minimum number of samples required at each leaf node, \texttt{min\_samples\_leaf} = 2. We evaluated \texttt{n\_estimators} from 0 to 900 and \texttt{min\_samples\_leaf} from 0 to 20, and although smaller \texttt{min\_samples\_leaf} values yielded lower RMSE, we adopted 2 instead of 1 to avoid overly fine partitioning. Each M31 halo star is passed through to the leaf nodes of the decision trees, where each leaf node contains two or more stars, and the resulting predicted [Fe/H] and [$\alpha$/Fe] values for each M31 halo star are obtained by averaging the predictions across all trees. The abundance ratio [$\alpha$/H] was defined as the sum of the estimated values of [Fe/H] and [$\alpha$/Fe].

\section{Results}\label{sec:results}
In this section, we present a model evaluation based on statistical metrics obtained from the test sample by applying the trained RF models. We also assess the contributions of the input features to the estimation.
We then examine and compare the distributions of iron and $\alpha$-element abundances in the synthetic data and the M31 halo on CMDs, color–color diagrams, and the R.A.-Decl. plane.
Here, we include only the samples with estimated $-0.7 \leq$ [$\alpha$/Fe] $\leq +0.9$ because RF predictions near the limits of the training set tend to cluster toward the boundary values. These samples have $-4.0 <$ [Fe/H] $< +0.26$. We present the results for 76,391 M31 halo giant candidates in this section.

\subsection{Model Evaluation} \label{sec:modeleval}
The performance of the model in estimating abundances was evaluated using the test sample through RMSE and $R^2$. 
    \begin{align}\label{eq:rmse}
        \text{RMSE} = \sqrt{\frac{1}{N} \sum_{i=1}^{N} (y_{{\rm phot},i} - y_i)^2}
    \end{align}
    
    \begin{align}\label{eq:r2}
        R^2 = 1 - \frac{\sum_{i=1}^{N} (y_{{\rm phot},i} - y_i)^2}{\sum_{i=1}^{N} (y_i - \bar{y})^2}
    \end{align}
where $N$ is the number of stars, $y_{{\rm phot},i}$ is the abundance estimate for the $i$th star, $y_i$ is the abundance of the corresponding synthetic star, and $\bar{y}$ is the mean of the synthetic abundances.

The RMSE values, which quantify how precise the predictions are compared to the synthetic abundance values, are 0.089 for [Fe/H] and 0.149 for [$\alpha$/Fe] in the test sample. Although the RMSE is somewhat larger for [$\alpha$/Fe], values of about 0.10 and 0.15 indicate relatively precise predictions. The $R^2$ values, which represent the relative magnitude of the prediction errors relative to the variance of the synthetic data, are 0.99 for [Fe/H] and 0.92 for [$\alpha$/Fe], indicating that in both cases the model reproduces the overall distribution of the synthetic data to a reasonable extent.

\begin{figure}[!t]
    \centering
    \hspace{-0.1in}
    \includegraphics[width=0.48\textwidth]{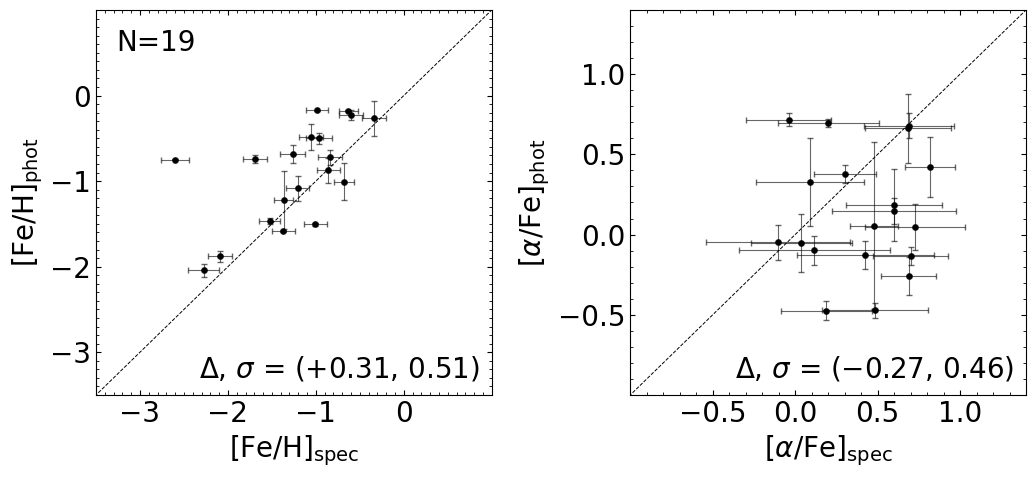}
    \caption{Comparison between DEIMOS spectroscopic measurements and M31 halo photometric estimates. The left and right panels show the comparisons for [Fe/H] and [$\alpha$/Fe], respectively. The number of overlapping samples is indicated in the upper left, and the residual $\Delta$ (phot $−$ spec) and the standard deviation $\sigma$ values are shown in the lower right.}
    \label{Fig:specphot}
\end{figure}

Another way to evaluate model performance is to compare it with spectroscopic data, as shown in Figure~\ref{Fig:specphot} using the DEIMOS sample. The left and right panels show [Fe/H] and [$\alpha$/Fe], respectively, and the dashed diagonal line indicates the one-to-one relation. While our photometric [$\alpha$/Fe] estimates use [Mg/Fe] as a proxy, the spectroscopic values represent the [$\alpha$/Fe] values adopted in the model atmospheres to determine individual elemental abundances, based on Mg, Si, Ca, and Ti absorption lines \citep{Kirby2020}.
In the lower right corner of each panel, the offset $\Delta$ (photometric $-$ spectroscopic) and the standard deviation $\sigma$ between the two are shown. 
After applying convex hull cuts to the M31 halo sample to bring it within the range of the synthetic photometry parameters, the cross-matched DEIMOS sample was reduced to 19 stars. Using this sample, we obtained mean ($\Delta$, $\sigma$) values of ($0.31, 0.51$) dex for [Fe/H] and ($-0.27, 0.46$) dex for [$\alpha$/Fe]. Thus, the photometric [Fe/H] estimates are higher, while the [$\alpha$/Fe] estimates are slightly lower than the spectroscopic values.

\begin{table}
\caption{Mean OOB Errors as a Function of $i_0$}
\resizebox{0.9\linewidth}{!}{
\hspace{-0.6cm}
\begin{tabular}{cccc}
\toprule\toprule
$i_0$ & Numbers & $\sigma_\mathrm{[Fe/H]_{mean}}$ & $\sigma_\mathrm{[\alpha/Fe]_{mean}}$ \\
{[mag]} & & [dex] & [dex] \\
\midrule
$[20.5,\ 21.5]$ & 4,491 & 0.11 & 0.18 \\
$[21.5,\ 22.0]$ & 12,060 & 0.10 & 0.16 \\
$[22.0,\ 22.5]$ & 20,307 & 0.10 & 0.16 \\
$[22.5,\ 23.0]$ & 23,491 & 0.08 & 0.15 \\
$[23.0,\ 23.5]$ & 13,697 & 0.06 & 0.12 \\
$[23.5,\ 24.5]$ & 2,345 & 0.04 & 0.08 \\
\bottomrule
\end{tabular}
}
\label{tab:i0_fehafeerr}
\end{table}

\begin{table}
\caption{Mean OOB Errors as a Function of $(g-i)_0$}
\resizebox{0.9\linewidth}{!}{
\hspace{-0.6cm}
\begin{tabular}{cccc}
\toprule\toprule
$(g-i)_0$ & Numbers & $\sigma_\mathrm{[Fe/H]_{mean}}$ & $\sigma_\mathrm{[\alpha/Fe]_{mean}}$ \\
{} & & [dex] & [dex] \\
\midrule
$[0.75,\ 1.00]$ & 2,247 & 0.04 & 0.06 \\
$[1.00,\ 1.25]$ & 12,163 & 0.04 & 0.07 \\
$[1.25,\ 1.50]$ & 28,469 & 0.09 & 0.18 \\
$[1.50,\ 1.75]$ & 26,369 & 0.10 & 0.17 \\
$[1.75,\ 2.05]$ & 7,143 & 0.08 & 0.12 \\
\bottomrule
\end{tabular}
}
\label{tab:gi0_fehafeerr}
\end{table}

\subsection{Estimation of Errors} \label{sec:error}
Following \citet{Lu2021}, we estimated the errors between the synthetic abundances and the RF estimates for each star using out-of-bag (OOB) samples, defined as the samples not included in the bootstrap training set for each tree. Specifically, in this study, the OOB samples are passed through the trained trees, and their estimated values are obtained. For each leaf involved in this process, the RMSE is computed and stored using the synthetic abundances and estimated abundances of the OOB samples. Subsequently, when the M31 halo samples are passed through the same trees to obtain their abundance estimates, the RMSE (derived from the OOB samples) of the leaf into which each sample falls is assigned as the error for that star. By averaging these errors over all trees, we defined the resulting value as the OOB error for each M31 halo star. The mean OOB error is 0.07 dex for [Fe/H] and 0.13 dex for [$\alpha$/Fe].

To further investigate whether the OOB error values depend on stellar brightness, we examined the mean OOB errors of M31 halo sample as a function of $i_0$, which are listed in Table~\ref{tab:i0_fehafeerr}. These values are calculated in 0.5 mag bins of $i_0$, and the number of M31 halo stars in each bin is also provided. The relatively small samples at the faint and bright ends are included at the edges of the magnitude range. From this, we found that the mean OOB errors decrease as $i_0$ increases, and that the error in [$\alpha$/Fe] is about 0.06 dex larger than that in [Fe/H].
Table~\ref{tab:gi0_fehafeerr} shows the mean OOB errors of the M31 halo sample as a function of $(g-i)_0$ in intervals of 0.25 mag. The errors are larger at $1.25 < (g-i)_0 \leq 1.75$ and decrease toward both bluer and redder colors, especially toward the blue side. This trend may be related to the distribution of the training set in the CMD and color-color diagram. At both extremes of $(g-i)_0$, the training set spans narrower ranges of [Fe/H] and [$\alpha$/Fe], causing the predictions to be within a narrower range and resulting in smaller OOB errors (see Section~\ref{sec:ccdcmd}).

\begin{figure}[!t]
    \centering
    \includegraphics[width=0.38\textwidth]{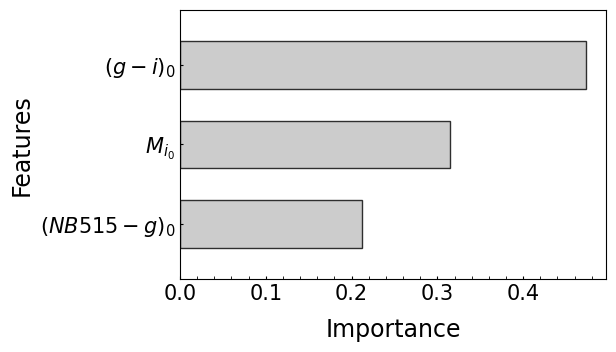}
    \caption{Feature importance, which shows the contribution of each input feature to the abundance estimation.}
    \label{Fig:feature_importances}
\end{figure}

\begin{figure}
    \centering
    \includegraphics[width=0.46\textwidth]{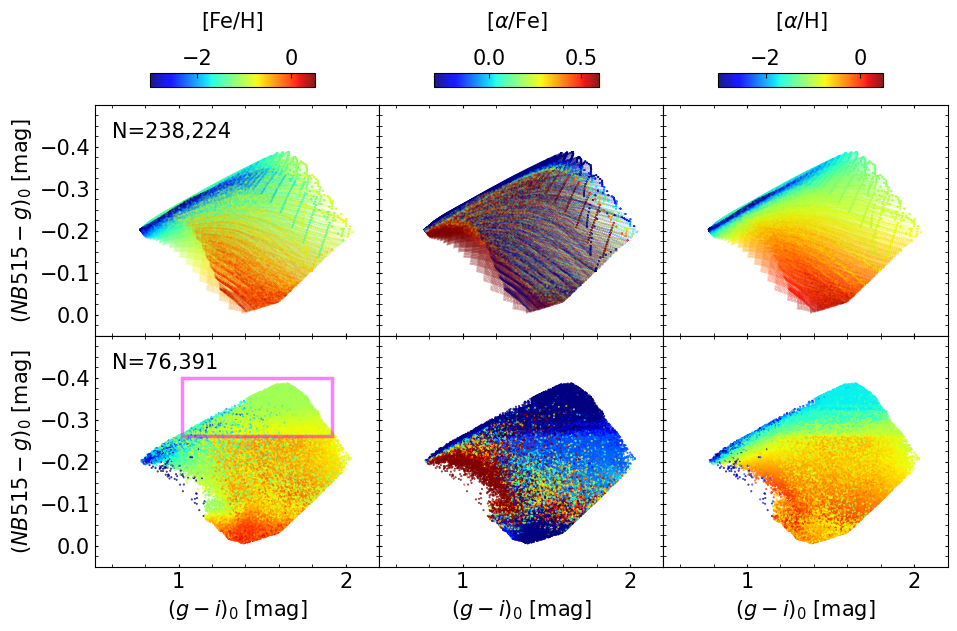}
    \includegraphics[width=0.455\textwidth]{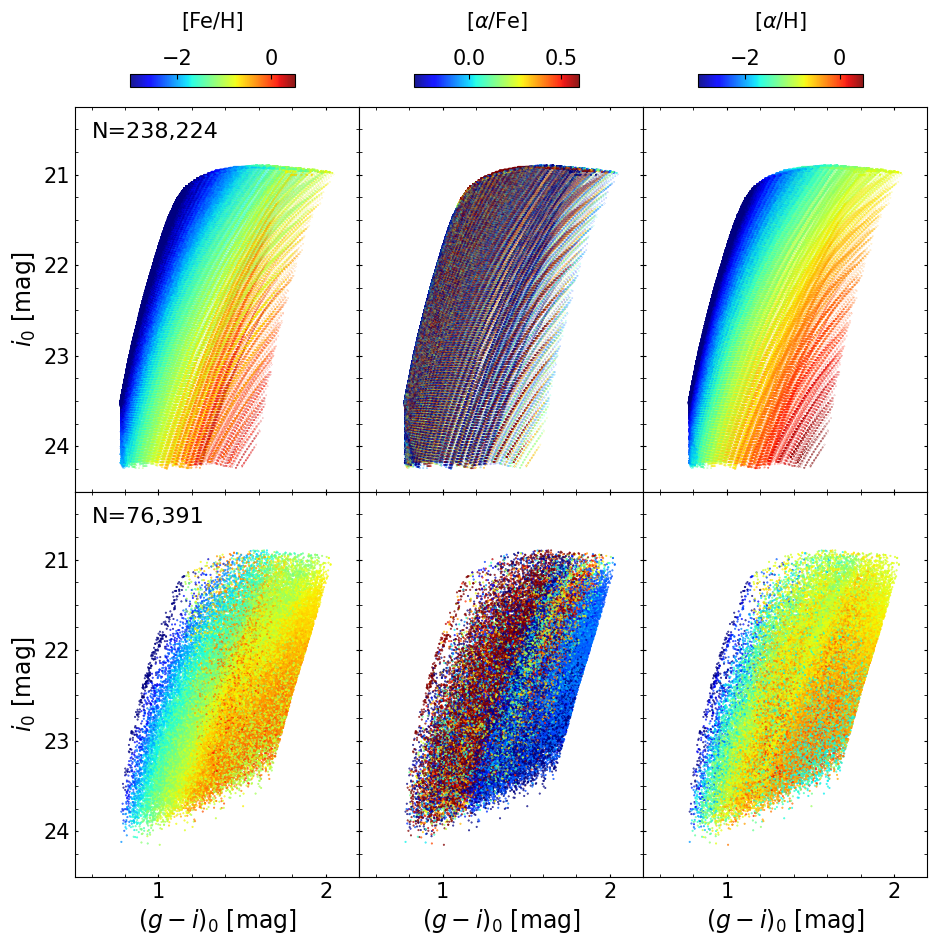}
    \centering
    \caption{Upper panels: Color-Color diagram of the synthetic photometric data used as a training set (top) and M31 halo sample (bottom), color-coded by [Fe/H], [$\alpha$/Fe], and [$\alpha$/H] from left to right. The number of samples in each panel is indicated in the upper left corner. The magenta-boxed region is discussed further in the text. Lower panels: Same as the upper panels, but for the CMDs.} 
    \label{Fig:cmdccd}
\end{figure}

\begin{table}
\caption{Mean Abundances and Standard Deviations}
\resizebox{0.9\linewidth}{!}{
\hspace{-0.3cm}
\begin{tabular}{ccc}
\toprule\toprule
Isochrone age & Abundance & M31 halo \\
\,[Gyr] &  & (mean$~\pm~$std) \\
\midrule
\multirow{3}{*}{7} 
& [Fe/H]        & $-0.76 \pm 0.46$ \\
& [$\alpha$/Fe] & $+0.01 \pm 0.42$ \\
& [$\alpha$/H]  & $-0.75 \pm 0.50$ \\
\midrule
\multirow{3}{*}{10} 
& [Fe/H]        & $-0.82 \pm 0.46$ \\
& [$\alpha$/Fe] & $+0.02 \pm 0.43$ \\
& [$\alpha$/H]  & $-0.80 \pm 0.49$ \\
\midrule
\multirow{3}{*}{13} 
& [Fe/H]        & $-0.85 \pm 0.48$ \\
& [$\alpha$/Fe] & $+0.04 \pm 0.43$ \\
& [$\alpha$/H]  & $-0.81 \pm 0.48$ \\
\bottomrule
\end{tabular}
}
\label{tab:abundance_summary}
\end{table}

\begin{figure*}
    \centering
    \includegraphics[width=0.95\textwidth]{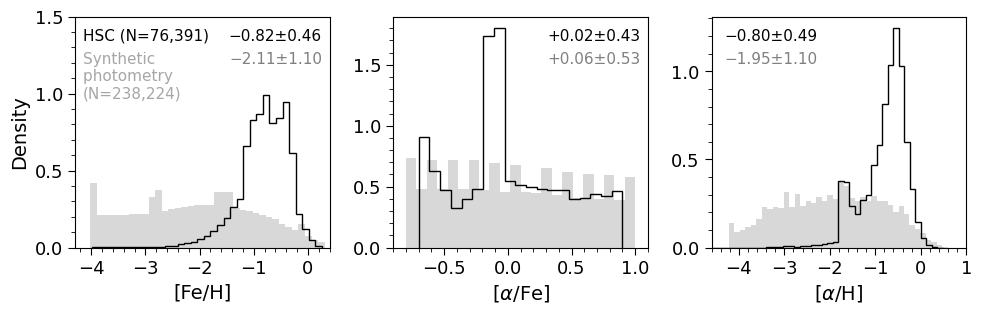}
    \centering
    \caption{Distribution of [Fe/H], [$\alpha$/Fe], and [$\alpha$/H]. The M31 halo and synthetic photometric samples are shown as black open and gray filled histograms, respectively. The sample sizes, mean abundances, and standard deviations are indicated at the top.}
    \label{fig:fehafeahhist}
\end{figure*}

\begin{figure}
    \centering
    \hspace{-0.15in}
    \includegraphics[width=0.42\textwidth]{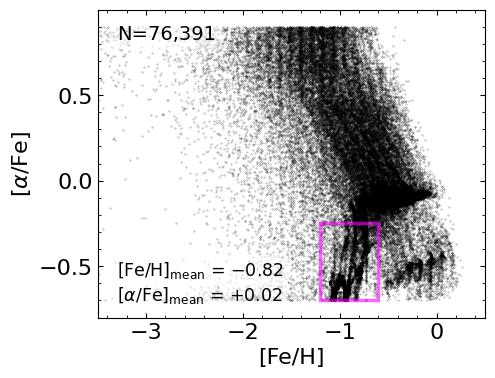}
    \centering
    \caption{Distribution of the [Fe/H] vs. [$\alpha$/Fe] estimates for the M31 halo sample. The sample within the magenta box appears to correspond to that within the magenta box in Figure~\ref{Fig:cmdccd}, as discussed further in the text.} 
    \label{fig:fehafedist}
\end{figure}

\subsection{Feature Importance} \label{sec:featureimportances}
The RF model quantifies the contribution of each input feature to the estimates based on the mean decrease in impurity (MDI)\@. Figure~\ref{Fig:feature_importances} shows the contribution of each input feature in predicting the [Fe/H] and [$\alpha$/Fe] abundances, where features with larger importance values, i.e., higher MDI scores, play a more significant role in the predictions.
As shown in the figure, the highest contribution of $(g-i)_0$ may reflect its role in tracing the abundance-dependent stellar loci in both the CMD and color-color diagram.
The higher importance of $M_{i_0}$ than $(NB515-g)_0$ may reflect its greater impurity reduction in the synthetic training data. For example, $M_{i_0}$ may better distinguish abundances than $(NB515-g)_0$ at a given $(g-i)_0$. In contrast, \citet{Hong2025} reported higher importance for NB515-related color indices and the lowest importance for $M_{i_0}$, possibly because their training data and input features differed from ours, as they used spectroscopic training data and six input features (three magnitudes and three colors).

\subsection{Color-Color and Color-Magnitude Diagram} \label{sec:ccdcmd}
The upper panels of Figure~\ref{Fig:cmdccd} show the color-color diagrams of the synthetic data and the M31 halo sample in the top and bottom rows, respectively. The abundance-dependent color trend from the upper left to the lower right is similarly reproduced in the photometric sample. 
However, a horizontal boundary appears in the photometric abundance estimates at $(NB515-g)_0 \approx -0.24$, with a relatively uniform color pattern above this boundary in the color-color diagram, as marked by the magenta box. Inspection of the $\alpha$-poor synthetic data showed that this feature roughly corresponds to a transitional change in the distribution of synthetic stars around $(g-i)_0 = 1.0$ and [Fe/H] $= -1$. At [Fe/H] $\lesssim -1$, the synthetic data largely overlap along a narrow linear-like locus, corresponding to the blue-to-cyan region in the [Fe/H]-coded upper left panel. At [Fe/H] $\gtrsim -1$, the sequences become more separated and curve toward redder $(g-i)_0$ colors. Therefore, this transition in the training data, combined with the tree-based splitting of the RF model, may produce the horizontal boundary and the limited color variation at $(NB515-g)_0 \leq -0.24$ in the photometric estimates.

The bottom panels of Figure~\ref{Fig:cmdccd} show the CMDs, and the [Fe/H]-coded panels also show broadly similar distributions between the synthetic and M31 halo samples. However, in the [$\alpha$/Fe]-coded CMD, a relatively distinct $\alpha$-rich population appears at lower $(g-i)_0$ side. As discussed in the caption to Figure~\ref{Fig:cmd_mist_chemfit}, this region corresponds to where the chemfit-derived colors for synthetic stars with [$\alpha$/Fe] $\leq -0.3$ become bluer than the original isochrones. This apparent concentration of $\alpha$-rich stars is also consistent with the lower-left region of the color-color diagram, where the most $\alpha$-rich isochrone occupies much of the region with little contribution from the other isochrones.

These color-color diagrams and CMDs are mainly used to examine whether the photometric abundance estimates broadly follow the abundance distributions of the synthetic photometric data. Also, given the properties of RF models, such as binary splitting that can produce discrete features, the narrow range of predicted values when the training-data range is limited, particularly in the extreme abundance regions, and our cuts on the most [$\alpha$/Fe]-poor and -rich estimates to reduce the accumulation of predictions near the boundaries, we focus more on relative abundance differences among the M31 halo components in the following analysis than on specific absolute abundance values.

\subsection{Photometric Abundance Distributions} \label{fehafedistribution}
Figure~\ref{fig:fehafeahhist} shows the distributions of [Fe/H], [$\alpha$/Fe], and [$\alpha$/H]\@. The black open histograms correspond to the photometric sample and the gray filled histograms show the synthetic photometric sample. The number of sample, mean abundances, and mean standard deviations of the abundance distributions are listed at the top of each panel, while the mean abundances estimated using isochrones of different ages are summarized in Table~\ref{tab:abundance_summary}.  From 7 to 10 to 13 Gyr, [Fe/H] decreases by 0.06 and 0.03 dex, while [$\alpha$/Fe] increases by 0.0006 and 0.03 dex, respectively.

The estimated [Fe/H] vs.\ [$\alpha$/Fe] distribution of the M31 halo sample is shown in Figure~\ref{fig:fehafedist}. The most noticeable features are the overall decreasing trend in [$\alpha$/Fe] as [Fe/H] increases, and a slightly clumped region around [$\alpha$/Fe] $= -0.1$. Using Figure~\ref{Fig:cmdccd}, we found that the stars in these features are distributed over $(NB515-g)>-0.24$ in the color-color diagram and over the full range in the CMD, although the clumped region is mainly found at $(g-i)_0>1.0$ in the CMD\@. 

A slightly increasing trend in [$\alpha$/Fe] with increasing [Fe/H] is also noticeable at [$\alpha$/Fe] $<-0.2$ and [Fe/H] $<-0.7$, as marked by the magenta box. We found that these stars preferentially occupy the $(NB515-g)<-0.24$ region, enclosed by the magenta box in the color-color diagram in Figure~\ref{Fig:cmdccd}, and the $(g-i)_0>1.0$ region in the CMD. As discussed in Section~\ref{sec:ccdcmd} for the color-color diagram, the synthetic data in this low $(NB515-g)_0$ region include more metal-poor stars and span a wider range of [$\alpha$/Fe], whereas the target estimates appear more concentrated at [Fe/H] $\gtrsim -1$ and toward the [$\alpha$/Fe]-poor end. Thus, this difference likely produced the apparent feature, which may not reflect a real chemical abundance trend in M31.

\begin{figure}
    \centering
    \includegraphics[width=0.45\textwidth]{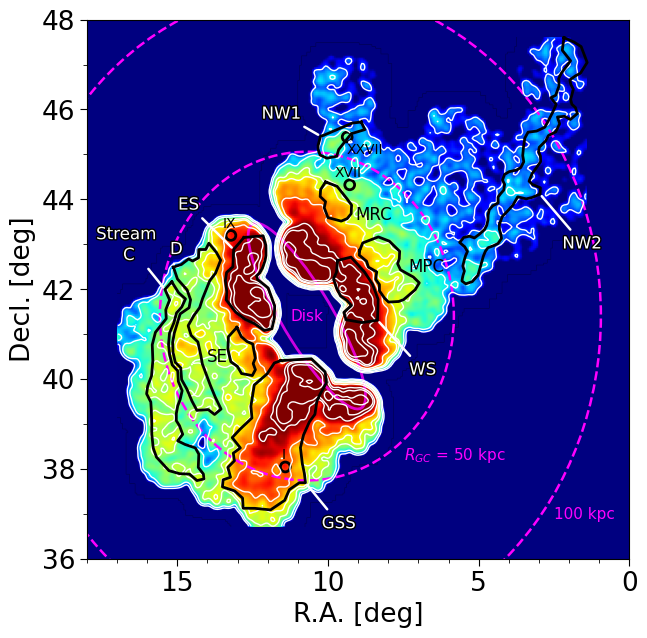}
    \centering
    \caption{Stellar number density contours showing the substructures of M31. The contour levels increase from 15\% to 95\% in steps of 10\% and are shown as white lines. Substructures in the southeast field, including GSS, Stream C and D, ES, and SE, and those in the northwest field, including NW1 and NW2 streams, MRC, MPC, and WS, are outlined with black lines. The And I, IX, XVII, and XXVII dSphs are marked with black circles. The disk region, defined from the radial surface brightness profile in \citet{Ogami2025}, is shown as a magenta solid ellipse at the center. The smooth halo is defined as the remaining stellar population after excluding stars associated with these substructures from the full sample. The ellipses at \Rgc\ = 50 and 100 kpc are indicated by magenta dotted ellipses.} 
    \label{fig:contour}
\end{figure}

\begin{figure*}
    \centering
    \hspace{0.3in}
    \includegraphics[width=0.8\textwidth]{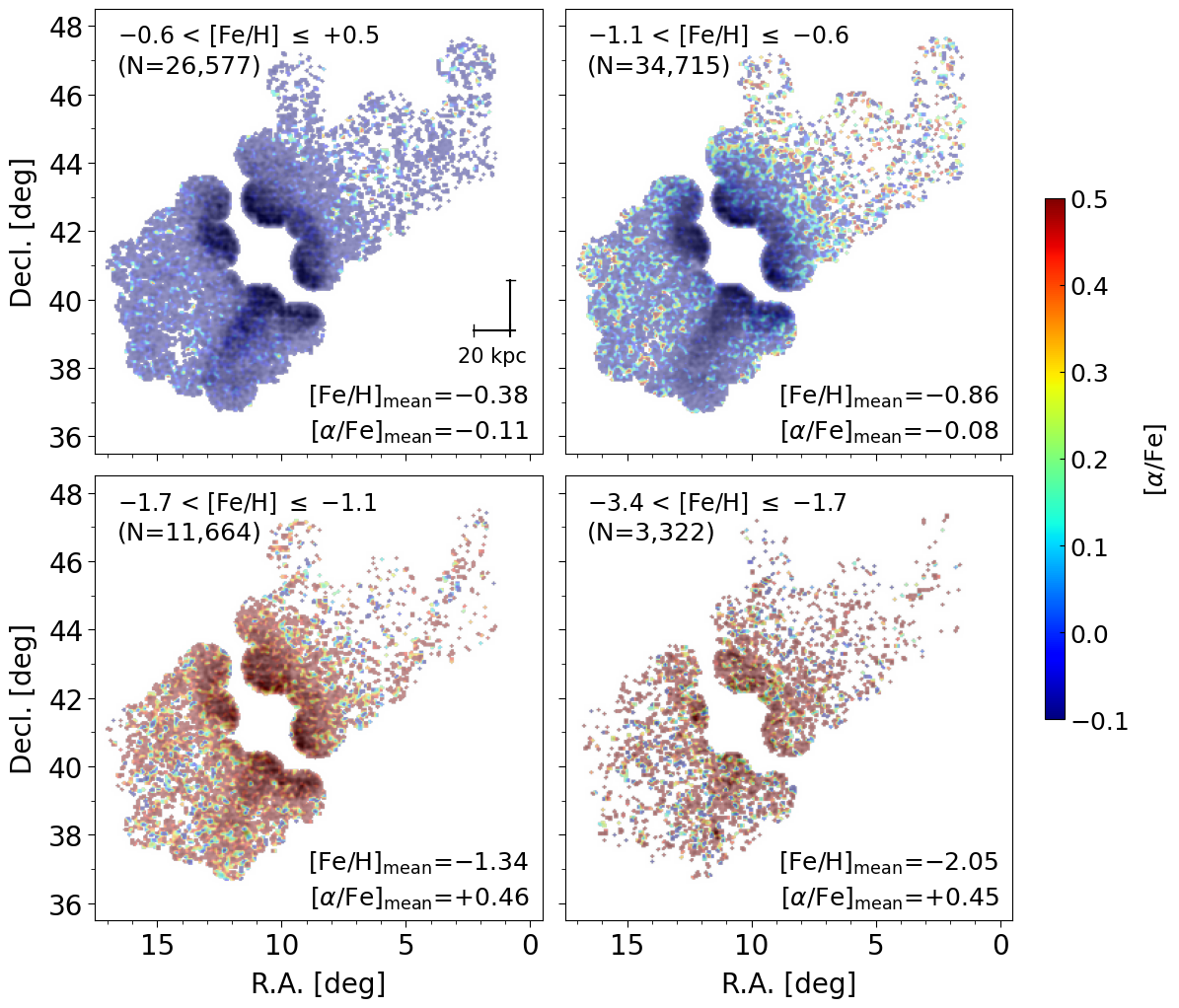}
    \caption{The [$\alpha$/Fe] distribution and stellar number density in each [Fe/H] bin. The panels are arranged in order of decreasing [Fe/H], from top to bottom and left to right. The [$\alpha$/Fe] abundance density is color-coded, and the stellar number density is overlaid in grayscale. A scale bar corresponding to 20 kpc is shown in the upper left panel. The number of stars in each metallicity bin is indicated in the upper left corner of each panel, and the mean abundance values are shown in the lower-right corner.}
    \label{fig:afe_numberdensity}
\end{figure*}

\begin{figure*}
    \centering
    \includegraphics[width=0.49\textwidth]{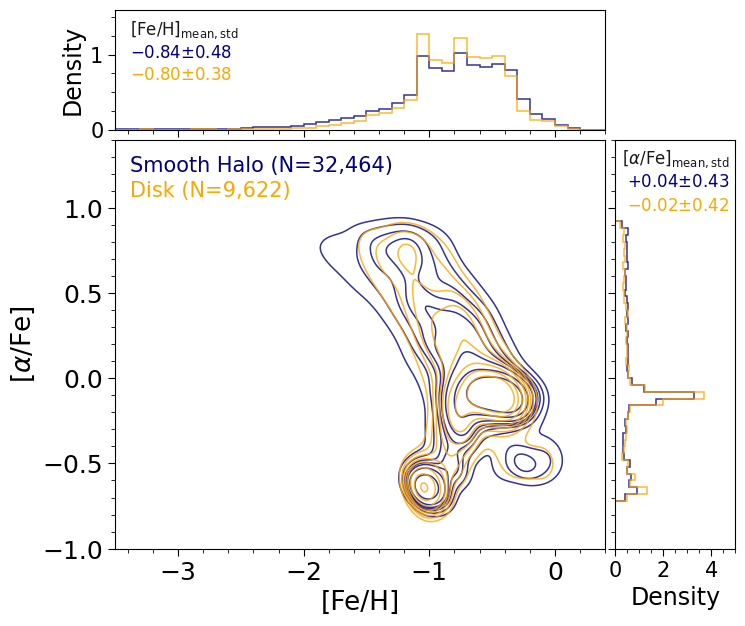}
    \includegraphics[width=0.49\textwidth]{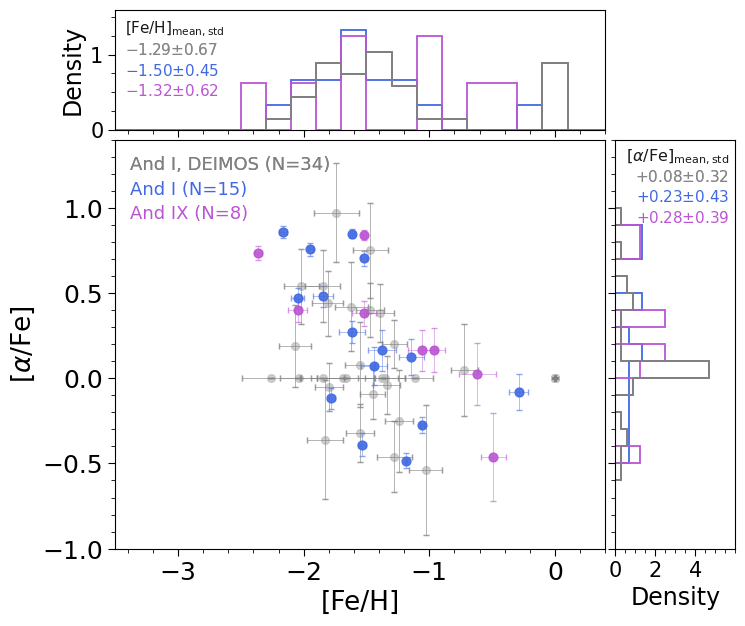}
    \includegraphics[width=0.49\textwidth]{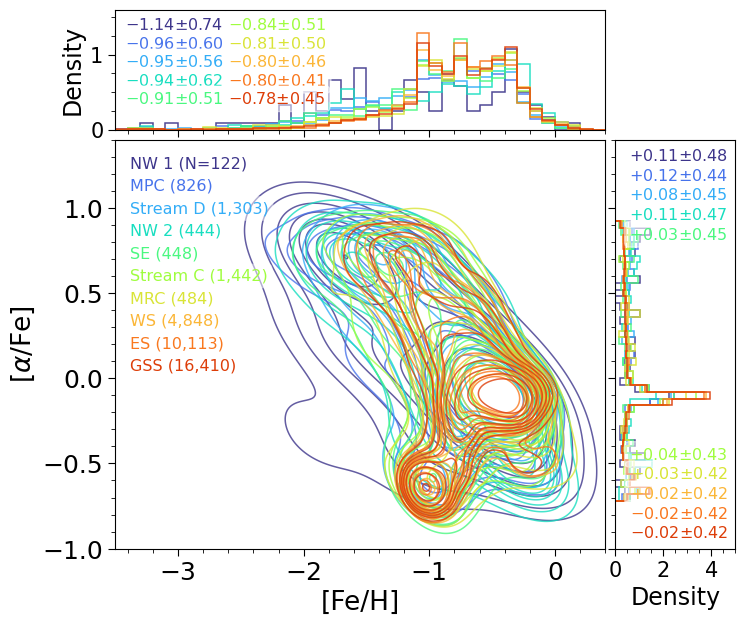}
    \caption{M31 substructures in the [Fe/H] vs.\ [$\alpha$/Fe] plane. Top left: Stars in the smooth halo and disk regions defined in this work, shown in navy and yellow, respectively. The number of stars in each component is listed in the upper left, and the mean and standard deviation values of the [Fe/H] and [$\alpha$/Fe] estimates are shown with the histograms in the top and right panels. Top right: Scatter distributions of the And I and IX dSphs, shown in blue and purple, respectively. Bottom: Same as the top left panel, but for all substructures identified in Figure~\ref{fig:contour}, except for the disk and dSphs, color-coded from dark purple to red in order of increasing mean $\rm [Fe/H]$.}
    \label{fig:subcontours}
\end{figure*}

\section{Discussion}\label{sec:discussion}
Here we examine the [$\alpha$/Fe] distributions of the M31 halo sample and explore [Fe/H] vs.\ [$\alpha$/Fe] across different components of the M31 halo and the smooth halo to infer their SFHs. We also compare our results with previous studies and the Milky Way halo.

\subsection{Spatial Abundance Distribution in M31} \label{sec:spatialdistribution}
Figure~\ref{fig:contour} shows different components of the M31 halo with contours at levels from 15\% to 95\% in steps of 10\%. Because our data set contains only photometric information for individual stars, we visually identified substructures from contour overdensities.

In addition to the structures identified by \citet{Ogami2025}, including the southeastern (SE) shell, metal-poor cloud (MPC), and metal-rich cloud (MRC), we also identified previously known substructures such as the giant stellar stream (GSS; \citealp{Ibata2001}), streams C and D (\citealp{Ibata2007}), eastern shelf (ES; \citealp{Ferguson2002}), western shelf (WS; \citealp{Fardal2007}), and NW streams 1 (NW1; \citealp{McConnachie2009, Richardson2011}) and 2 (NW2; \citealp{McConnachie2009}), outlined with black solid lines. 
The dwarf spheroidal galaxies (dSphs) within the observation field, And I, IX, XVII, and XXVII are marked with black circles. The positions, half-light radii, ellipticities, position angles of the major axes measured from east to north, and distances of the dSphs are adopted from \citet{McConnachie2012} and the references therein \citep{McConnachie2006, Collins2010, Collins2011, Brasseur2011, Richardson2011}.

Additionally, the disk is indicated by a solid magenta ellipse near the center. This ellipse was defined by adopting 2.5 exponential scale lengths of the radial surface brightness profile, $R_d = 14.99$ kpc, measured by \citet{Ogami2025}. This disk region also includes the location at \Rgc\ = 26 kpc along the major axis, studied as the outer disk by \citet{Escala2020a}. We adopt an ellipticity of 0.73 \citep{Courteau2011, Dalcanton2023} and a position angle of $43^\circ$ \citep{Dalcanton2012}. Finally, we define the smooth halo of M31 as the complete sample with all the aforementioned substructures removed.
The ES and GSS substructures spatially overlap with the region defined as the disk in this work, and we found that about 13\% and 5\% of their stars are in common, respectively. We also define the final GSS sample by excluding And I stars located within the identified GSS region.

Figure~\ref{fig:afe_numberdensity} shows the [$\alpha$/Fe] distribution and the stellar number density in each metallicity bin of the full giant sample. The [$\alpha$/Fe] is color-coded, while the stellar number density is overlaid as a grayscale map, ranging from low-density (transparent) to high-density (black), both smoothed using a Gaussian filter in \texttt{SciPy} with $\sigma = 1$ deg, where $\sigma$ represents the standard deviation of the Gaussian kernel. The [Fe/H] ranges adopted for the four panels are based on \citet{Ibata2014}, with the most metal-rich and metal-poor bins slightly extended beyond their boundaries. 

In the upper left panel ($-0.6 < \mathrm{[Fe/H]} \leq +0.5$), a large number of stars are found near the disk or inner halo regions, as well as in the GSS\@, and are predominantly shown in darker blue colors, indicating relatively low [$\alpha$/Fe] values. In the upper right panel ($-1.1 < \mathrm{[Fe/H]} \leq -0.6$), the stellar number density near the disk and in the southeast region decreases somewhat and becomes greener, indicating moderately higher [$\alpha$/Fe]. In the lower-left panel ($-1.7 < \mathrm{[Fe/H]} \leq -1.1$), stars exhibit much redder colors, indicating higher [$\alpha$/Fe], and the stellar number density in the southeast region appears to decrease. In the northwestern region, NW1 and NW2 stream look more prominent. In the lower-right panel ($-3.4 < \mathrm{[Fe/H]} \leq -1.7$), corresponding to the most metal-poor stars, the overall stellar number density further decreases in both regions. Nevertheless, some stellar concentrations remain near the disk, GSS, And I, and the northern stream regions, and a smoothly distributed halo component is visible across the field.

Compared with \citet{Ibata2014}, we found broadly consistent behavior in that the GSS is detected most prominently in the $-0.6 < \mathrm{[Fe/H]} \leq +0.5$ bins. The NW streams also appear more prominently at $-1.7 < \mathrm{[Fe/H]} \leq -1.1$, although NW2 is also visible in our more metal-rich bin of $-1.1 < \mathrm{[Fe/H]} \leq -0.6$, which does not appear to be the case in \citet{Ibata2014}. The most notable difference is that we were also able to capture the increase in [$\alpha$/Fe] with decreasing [Fe/H] using the $NB515$ filter; it contributes not only to the estimation of $\alpha$ abundances but also to differences in giant members selection. There are also slight differences in the samples; for example, the PAndAS CFHT/MegaCam dataset of \citet{Ibata2014} included stars with $i_0 < 23.5$ mag and $0.8 < (g - i)_0 < 1.8$, whereas our sample extends to fainter and redder stars ($i_0 < 24.2$ mag and $0.8 < (g - i)_0 < 2.0$). In addition, the adopted models and analysis methods may also play a role in this discrepancy, such as the use of Dartmouth 13 Gyr isochrones in \citet{Ibata2014} compared with MIST 10 Gyr isochrones and an RF-based approach in this work (see Section~\ref{sec:radialgradients} for further discussion).

\subsection{Abundance Distribution of M31 Substructures} \label{sec:fehafe_substructures}
The [Fe/H] vs.\ [$\alpha$/Fe] distributions of the identified substructures and the smooth halo can be seen in Figure~\ref{fig:subcontours}. 
The mean and standard deviation values of [Fe/H] and [$\alpha$/Fe] estimates are provided with the histograms at the top and right. The upper left panel shows stars in the smooth halo and disk regions, shown in navy and yellow, respectively. The smooth halo contains about 32,500 stars and shows a slightly more metal-poor and $\alpha$-enhanced population than the disk. Thus, we infer that the progenitors of the M31 halo may have formed at earlier epochs, resulting in an older stellar population, while the outer disk likely experienced the formation of relatively younger stars until more recent epochs.

The upper-right panel shows the And I and And IX dSphs. And XVII and And XXVII are not included because only one and zero stars, respectively, are identified. Compared with the smooth halo in the top left panel, the dSphs show more metal-poor and $\alpha$-rich distributions. Compared with the DEIMOS spectroscopic measurements of And I from \citet{Kirby2020}, shown in gray, the photometric estimates in blue are more metal-poor and $\alpha$-rich. Compared with the co-added DEIMOS results of \citet{Wojno2020}, with mean [Fe/H] values of $-1.36$ and $-2.03$ for And I and And IX, respectively, our [Fe/H] estimate is lower for And I but higher for And IX.
Comparing the relative abundances of And I and IX, \citet{Kirby2013} and \citet{Wojno2020} found that the less massive And IX has a lower mean [Fe/H] than And I, consistent with the mass--metallicity relation. However, in our photometric estimates, the two galaxies span nearly the same range in [$\alpha$/Fe], and And IX extends to higher [Fe/H] when considering the standard deviations. This may partly arise from differences in dSph member selection: \citet{Kirby2013} and \citet{Wojno2020} selected dSph members using spatial and CMD positions as well as spectroscopic information, whereas our photometric member identification was mainly based on visual inspection of their spatial distributions, which may have resulted in contamination of the And IX sample by relatively metal-rich M31 outer disk stars. Future observations targeting M31 dSphs including their central and outskirt regions will increase the member sample size and enable a more precise comparison of their chemical properties.

The bottom panel shows all substructures except the smooth halo, disk, and dSphs. The contours are colored from dark purple to red in order of increasing mean [Fe/H], with the substructure names labeled accordingly, and the top and right histograms show the mean abundances and standard deviations. From the most metal-poor to the most metal-rich, the substructures are NW1, MPC, Stream D, NW2, SE, Stream C, MRC, WS, ES, and the GSS\@. Overall, [$\alpha$/Fe] tends to increase toward lower [Fe/H]. One interesting result from this relative abundance comparison is that the GSS, which contains the largest number of stars among all substructures, has the highest mean [Fe/H] and the lowest mean [$\alpha$/Fe]. This is consistent with expectations that the massive GSS structure may have experienced a more complex and extended SFH. 
Additionally, compared with the smooth halo in the top left panel, the contours of metal-rich substructures, including the GSS, show similar distributions. This similarity may partly result from the visual identification of the components; for example, the stars identified as belonging to the GSS are likely to include a non-negligible fraction of smooth-halo stars. Such smooth-halo contamination within the identified components may therefore contribute to the similar distributions.

\subsection{Radial Abundance Gradients} \label{sec:radialgradients}

\begin{figure*}
    \centering
    \includegraphics[width=0.39\textwidth]{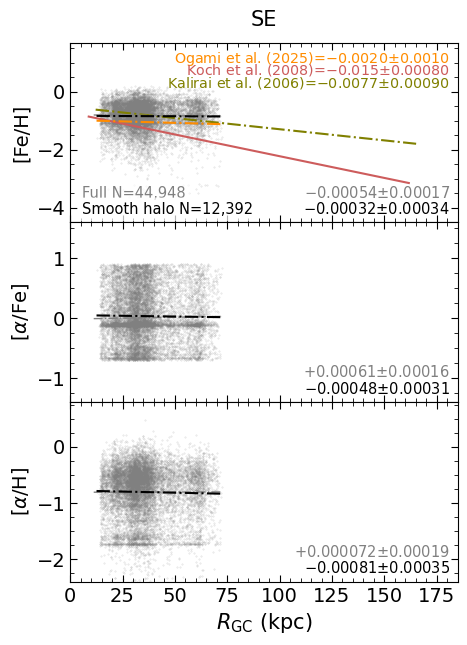}
    \includegraphics[width=0.39\textwidth]{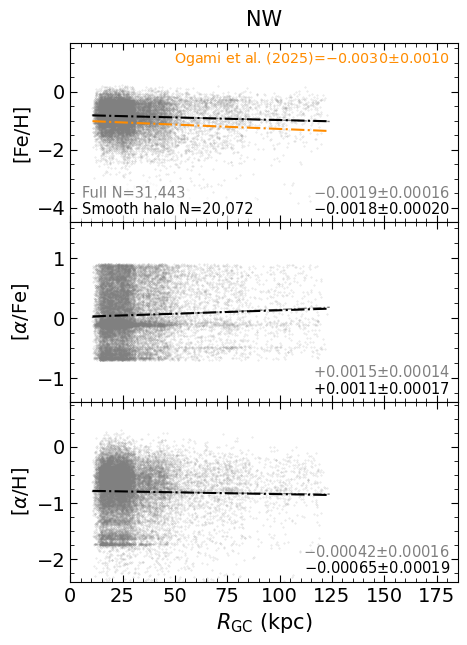}
    \includegraphics[width=0.39\textwidth]{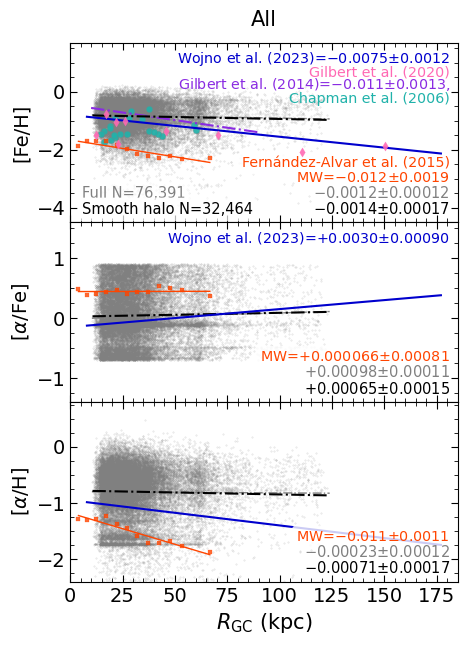}
    \caption{Top left: Radial abundance gradients of the M31 smooth halo sample in the southeast region. From top to bottom, the panels show the gradients of [Fe/H], [$\alpha$/Fe], and [$\alpha$/H], respectively. Black dash-dot lines represent the least-squares linear fits, with their slopes and fitting uncertainties shown in the lower right corner. Gray represents the fit results for the full M31 halo sample. 
    Yellow, brown, and olive indicate the results from \citet{Ogami2025}, \citet{Koch2008}, and \citet{Kalirai2006}, respectively. Studies using photometric and spectroscopic data are shown as dashed-dotted and solid lines, respectively. 
    Top right: Same as the left, but for the northwest region, with the results from \citet{Ogami2025}. 
    Bottom: Same as the top left, but for the entire region. Blue, pink, purple, and turquoise represent the results from \citet{Wojno2023}, \citet{Gilbert2020}, \citet{Gilbert2014}, and \citet{Chapman2006}, respectively. Orange represents radial gradients of the Milky Way halo derived from low-resolution SDSS/SEGUE spectra \citep{Fernandez-Alvar2015}.}
    \label{fig:radialgradients}
\end{figure*}

The left panel of Figure~\ref{fig:radialgradients} shows the radial gradients of [Fe/H], [$\alpha$/Fe], and [$\alpha$/H], from top to bottom, for the M31 smooth halo stars shown as gray dots in the southeast region over $12.6 < R_{GC} < 72.3$ kpc. The least-squares linear fit is shown as a black dash dot line, and its slope and standard deviation values are listed in the lower right in black. The gray text indicates the fit results for the entire M31 halo sample.

In the top panel, the smooth halo stars in the southeastern region show a negative [Fe/H] gradient toward larger radii. This negative slope is consistent with previous studies, including \citet{Kalirai2006}, \citet{Koch2008}, and \citet{Ogami2025}, shown in olive, brown, and yellow, respectively. The dash dot and solid lines represent results from photometric and spectroscopic studies, respectively.

\citet{Kalirai2006} estimated photometric [Fe/H] for 261 RGB stars at \Rgc\ $=12$ to 165 kpc using Kitt Peak National Observatory (KPNO) and CFHT photometry with 12 Gyr, [$\alpha$/Fe] $=0.0$ isochrones from \citet{VandenBerg2006}, and their gradient value here is adopted from \citet{Koch2008}. \citet{Koch2008} analyzed 1,316 RGB stars at \Rgc\ $=9$ to 160 kpc using DEIMOS spectroscopy and found a steeper gradient. They noted that the overestimation of photometric metallicities in \citet{Kalirai2006} may arise from uncertainties in the adopted isochrones and age-metallicity degeneracies, and may be particularly significant for metal-poor stars ([Fe/H] $<-1.5$), for which metallicity differences are difficult to discern from colors in the bluer region of the CMD.

\citet{Ogami2025}, using the same photometric dataset as ours, adopted 13 Gyr, [$\alpha$/Fe] $=0.0$, $-2.5<$ [Fe/H] $<0.5$ Dartmouth isochrones with $(g-i)_0<2.5$ and $i_0<23$, and found an [Fe/H] gradient shallower than those of \citet{Kalirai2006} and \citet{Koch2008}, but still steeper than ours. Similarly, their [Fe/H] gradient in the northwestern region, shown in the top right panel of Figure~\ref{fig:radialgradients}, is steeper than ours. In comparison, our MIST isochrones adopt an age of 10 Gyr, $-0.7<$ [$\alpha$/Fe] $<+0.9$, and $-2.1<$ [Fe/H] $<1.1$, with $0.8<(g-i)_0<2.0$ and $i_0<24.2$. 

When we used 13 Gyr MIST isochrones, we obtained an even flatter [Fe/H] gradient of $-0.000027\pm0.00036$ dex kpc$^{-1}$ and $-0.0019\pm0.00010$ dex kpc$^{-1}$ for the southeastern and northwestern regions, respectively, so the gradient differences between previous studies and this study may result from a combination of giant member selection (\citealt{Kalirai2006} and \citealt{Koch2008} used spectroscopically selected RGB samples, whereas \citealt{Ogami2025} used the color-color diagram with RGB probabilities and the Galactic-latitude dependence of foreground dwarf stars), magnitude and color cuts, stellar models, adopted [$\alpha$/Fe] ranges, and abundance determination methods, including photometric metallicity estimates for metal-poor stars, as discussed by \citet{Koch2008}. Spectroscopic samples can also be affected by field-to-field target selection and spatial sampling.

Considering the [$\alpha$/Fe] gradients in the middle panels of the two upper plots in Figure~\ref{fig:radialgradients} together with the [Fe/H] gradients, we found that removing substructures made the negative [Fe/H] gradients slightly flatter and the positive [$\alpha$/Fe] gradients flatter as well, with the southeastern region even having a negative [$\alpha$/Fe] slope. This suggests that including substructures enhances the abundance contrast across the halo. We found that this abundance contrast is larger in the inner halo in the southeastern region, but larger in the outer halo in the northwestern region. Indeed, the substructure fractions in 5 kpc radial bins exceed 50\% at \Rgc\ $\leq30$ kpc and $40\leq R_{GC}\leq60$ kpc in the southeastern halo. In the northwestern region, they reach about 35\% at $55\leq R_{GC}\leq60$ kpc and $85\leq R_{GC}\leq95$ kpc. In addition, the smooth halo shows lower mean [Fe/H] and higher mean [$\alpha$/Fe] than the substructures in both regions. Therefore, removing substructures produces larger decreases in [Fe/H] and increases in [$\alpha$/Fe] where they contribute more strongly, but smaller changes elsewhere, resulting in different gradients between the full and smooth-halo samples. Furthermore, the overall tendency of the smooth halo to be more [Fe/H] poor and [$\alpha$/Fe] rich than the substructures may suggest that it was accreted at an early stage, before its progenitor galaxies had experienced substantial enrichment from Type~Ia supernovae.

The bottom panels in Figure~\ref{fig:radialgradients} show the abundance gradients for the entire observed region. The overall [Fe/H] gradient is negative, consistent with previous studies including \citet{Ibata2014}, whereas [$\alpha$/Fe] shows a positive gradient. Before removing substructures, the southeastern and northwestern regions account for about 59\% and 41\% of the full sample, respectively; after removal, these fractions change to about 38\% and 62\%. Additionally, the northwestern sample extends over \Rgc\ $= 10.7$ to 122.5 kpc, much farther than the southeastern sample, so beyond about 72 kpc the combined result is determined only by the northwestern stars. Hence, the larger contribution and wider radial coverage of the northwestern region lead to decreasing [Fe/H] and increasing [$\alpha$/Fe] toward the outer halo.

The previous studies compared with our results for the entire smooth halo sample are as follows: \citet{Chapman2006}, shown as turquoise circles, measured metallicities from DEIMOS stacked spectra of 827 kinematically selected non-rotating RGB halo stars at \Rgc\ $=10$ to 70 kpc and reported no clear gradient, with a mean [Fe/H] of about $-1.4$. \citet{Gilbert2014} estimated photometric [Fe/H] for 1,549 M31 RGB stars using 10 Gyr, [$\alpha$/Fe] $=0.0$ isochrones \citep{VandenBerg2006} with $-2.3 \leq {\rm [Fe/H]} \leq 0.5$. They noted that, if rotating stars are present in the halo, the velocity selection of \citet{Chapman2006} may remove them, possibly contributing to the lack of a clear gradient. After removing tidal debris, \citet{Gilbert2014} derived a clear negative [Fe/H] gradient of $-0.0105 \pm 0.0013$ dex~kpc$^{-1}$ from the field median [Fe/H] at \Rgc\ $=9$ to 90 kpc, shown here as the purple dash dot line.

\citet{Gilbert2020} obtained 23 [Fe/H] and nine [$\alpha$/Fe] spectral-synthesis measurements from DEIMOS spectra of halo stars around M31 dSphs at \Rgc\ $=43$ to 165 kpc, supporting the negative [Fe/H] gradient found by \citet{Gilbert2014}. However, they did not report a specific slope, noting that a larger sample is needed to improve the sparse field-to-field spectroscopic sampling and reduce the selection bias caused by excluding metal-rich stars with strong TiO absorption that could not be reliably modeled. Therefore, we show only their measurements here as pink diamonds. \citet{Gilbert2020} also explained that the assumption of a single age and [$\alpha$/Fe] in estimating photometric [Fe/H] in \citet{Gilbert2014} may partly account for differences from their spectroscopic results, and the single stellar age adopted in this work may similarly contribute to the difference in gradient slope relative to the spectroscopic result.

The blue solid line shows the result from \citet{Wojno2023}, who measured the gradient over \Rgc\ $=8$ to 177 kpc using DEIMOS data and a spectral co-adding method, with 155 abundance measurements. In particular, by including distant metal-poor stars at \Rgc\ $>50$ kpc through co-addition, they obtained a clear negative [Fe/H] gradient. As shown in the middle panel, they also measured an [$\alpha$/Fe] gradient that increases outward, similar to our result, although their gradient is steeper. The difference in gradient slope may partly arise from differences between spectroscopic and photometric abundance measurements, sample selection, and radial coverage, as discussed above and by \citet{Wojno2023}.

As a result, the radial abundance gradients commonly found in previous studies and in this work allow us to broadly infer the assembly history of the entire M31 smooth halo: the inner region is relatively [Fe/H]-rich and [$\alpha$/Fe]-poor, likely reflecting larger contributions from more metal-rich accreted and in situ populations, whereas the outer region becomes [Fe/H]-poor and [$\alpha$/Fe]-rich, possibly due to a larger contribution from earlier accreted components \citep{Wojno2023}. However, more detailed star formation histories require accounting for factors such as the non-uniform radial distribution of substructures in the southeastern and northwestern regions and stellar age variations with radius, as well as more accurate abundance measurements over a wider area of the M31 halo.

\subsubsection{Comparison with the Milky Way Halo} \label{sec:m31mw}
Furthermore, we compared the radial abundance gradients of the M31 smooth halo with those of the Milky Way. \citet{Fernandez-Alvar2015} examined the distributions of median [Fe/H] and [Mg/H] as a function of Galactocentric distance \rgc\ for Milky Way halo stars located at $|Z|>5$ kpc. Using their sample of 2,835 stars with low-resolution SDSS/SEGUE DR10 spectra \citep{Ahn2014}, we applied the same Galactic height and [Fe/H] error cuts and calculated a projected Galactocentric radius for each star. Then, we rebinned the Milky Way sample in 5 kpc intervals in projected radius following their binning method and derived the median [Fe/H], [Mg/H], and [Mg/Fe]. We derived [Mg/Fe] from their [Mg/H] and [Fe/H] measurements and used it as a proxy for [$\alpha$/Fe] here, although a more direct comparison would require considering multiple $\alpha$-elements. 

Although the sample of \citet{Fernandez-Alvar2015} has more limited radial coverage than the M31 smooth halo sample, both show decreasing [Fe/H] toward larger radii. However, [Fe/H] decreases much more rapidly in the Milky Way, whereas relatively high [Fe/H] values are maintained to larger radii in the M31 smooth halo. \citet{Fernandez-Alvar2015} interpreted the radial metallicity variation in the Milky Way as resulting from larger contributions of fragments from more massive progenitor galaxies in the inner halo and from lower-mass and lower-metallicity accreted systems in the outer halo. From this perspective, the flatter [Fe/H] gradient in M31 may indicate that the contrast in progenitor mass and chemical enrichment between the inner and outer halo is smaller than in the Milky Way. Alternatively, relatively massive and chemically enriched progenitors may have contributed more strongly, with their debris extending to larger radii in the M31 smooth halo \citep[e.g.,][]{DSouza2018b, Gilbert2020}.

In contrast, the M31 smooth halo shows a stronger positive [$\alpha$/Fe] gradient than the Milky Way. If this difference is intrinsic, it may reflect different radial distributions of the star formation timescales and histories of the progenitors that built the two halos. However, the M31 [$\alpha$/Fe] estimates and Milky Way [Mg/Fe] measurements are based on different abundance definitions and methods, and the Milky Way outer-halo [Mg/Fe] measurements also show substantial scatter \citep{Fernandez-Alvar2015}. Therefore, caution is required in directly interpreting this difference as a systematic difference in the star formation or accretion histories of the two galaxies. In the future, larger spectroscopic samples covering comparable radial ranges in both galaxies, particularly extending farther into their outer halos, are needed to better characterize the progenitors and their star formation histories, thereby improving our understanding of the chemical enrichment and assembly histories of the two large galaxies in the Local Group.

\section{Summary}\label{sec:summary}
We used the Fe and $\alpha$ abundances estimated from photometric data for about 76,400 stars observed with the PAndAS broad-band $g$ and $i$ filters and the Subaru/HSC $NB515$ narrow-band filter to investigate the chemical abundance distributions and SFHs of the halo components of M31. The giant members were selected using polygon selection on the $gi$ color-magnitude diagram and the color-color diagram using the $NB515$ magnitude \citep{Komiyama2018b, Okamoto2024, Ogami2025, Hong2025}. This narrow-band filter targets Mg absorption features that are sensitive to surface gravity, making it effective not only for estimating $\alpha$-element abundances such as Mg, but also for distinguishing M31 giant stars with low surface gravity from foreground Milky Way dwarf stars. To construct the synthetic photometric data set used for model training, we obtained 7, 10, and 13 Gyr MIST isochrones with [$\alpha$/Fe] $= -0.2, 0.0, +0.2, +0.4$, and $+0.6$, and restricted them to the parameter range of the M31 halo giant sample. We derived the $(NB515 - g)_0$ and [$\alpha$/Fe] values of the isochrones using Chemfit, and obtained the final synthetic photometric data set by a factor of three using DFInterpolator with abundances, EEP, and ConvexHull on color-color diagram and CMD. 

Elemental abundances were estimated using an RF regression model. The synthetic photometric sample, with $(g-i)_0$, $M_{i_0}$, and $(NB515-g)_0$ as input features and [Fe/H] and [$\alpha$/Fe] as target values, was used to train the model and estimate [Fe/H] and [$\alpha$/Fe] for the M31 halo giant sample. To avoid clustering of predictions near the boundaries of the training set, we restricted the estimates to $-0.7 \leq [\alpha/{\rm Fe}] \leq +0.9$. We also estimated errors using OOB samples, obtaining mean errors of 0.07 dex for [Fe/H] and 0.13 dex for [$\alpha$/Fe]. The mean OOB errors decrease toward fainter $i_0$, with [$\alpha$/Fe] errors overall about 0.06 dex larger than [Fe/H] errors, and are largest at $1.25 < (g-i)_0 \leq 1.75$ but decrease toward both bluer and redder colors, likely because the training set spans narrower abundance ranges at the color extremes. The RF feature importances showed that $(g-i)_0$ contributed the most, followed by $M_{i_0}$ and $(NB515-g)_0$.

For the results, we presented the 10 Gyr isochrone case as an example. The M31 sample broadly follows the abundance distributions of the synthetic photometric data in both the color-color diagrams and CMDs. A noticeable boundary around $(NB515-g)_0=-0.24$ and $(g-i)_0=1.0$ likely arises from the narrow training-data locus at [Fe/H] $\lesssim -1$ and RF binary splitting, while the $\alpha$-rich population at lower $(g-i)_0$ in the CMD is related to the bluer chemfit-derived colors of synthetic stars with [$\alpha$/Fe] $\leq -0.3$. We obtained mean [Fe/H] and [$\alpha$/Fe] values of $-0.82$ and $+0.02$, respectively.  Choosing a different age would shift all the abundances in one direction.  As a result, we focused on the relative abundance distributions among the M31 halo components rather than on the absolute predicted values.

Using a stellar number density map of the M31 halo in equatorial coordinates, we identified several components, including streams, dSphs, the disk, and the smooth halo. The [$\alpha$/Fe] distributions in [Fe/H] bins on the R.A.-Decl.\ plane showed that the GSS and NW streams appear in similar metallicity bins to those found by \citet{Ibata2014}, although NW2 is also visible in our more metal-rich bin. More importantly, the use of the $NB515$ filter allowed us to detect increasing [$\alpha$/Fe] toward lower [Fe/H]. From the [Fe/H] vs.\ [$\alpha$/Fe] distributions, we found that the $\sim$32,500 smooth halo stars are slightly more metal-poor and $\alpha$-enhanced than the disk stars, possibly indicating earlier formation of the M31 halo progenitors and an older stellar population. We also examined the abundance distributions of And I and IX by comparison with previous studies, but future observations covering their central and outer regions are needed to increase the samples and improve comparisons of their chemical properties. Among the other substructures, the GSS has the relatively highest mean [Fe/H] and lowest mean [$\alpha$/Fe], consistent with the expectation that this massive structure experienced a more complex and extended SFH.

Examining the radial abundance gradients with least-squares linear fitting for the M31 smooth halo stars, we found decreasing [Fe/H] and generally increasing [$\alpha$/Fe] toward the outer halo, although the southeastern and northwestern regions show slightly different trends, consistent with previous studies \citep{Kalirai2006, Chapman2006, Koch2008, Gilbert2014, Gilbert2020, Wojno2023, Ogami2025}. However, our [Fe/H] estimates are generally higher, and the abundance gradients are flatter, which may arise from differences in sample selection, radial coverage, stellar models and single-age assumptions, and photometric and spectroscopic abundance measurements. In addition, removing substructures generally makes the [Fe/H] and [$\alpha$/Fe] gradients flatter, and the smooth halo is more metal-poor and $\alpha$-rich than the substructures, possibly indicating earlier accretion before substantial Type~Ia supernova enrichment. 

Compared with the SDSS/SEGUE Milky Way halo sample of \citet{Fernandez-Alvar2015}, the halos of both galaxies show negative [Fe/H] and positive [$\alpha$/Fe] radial trends. However, relatively high [Fe/H] values are maintained to larger radii in the M31 smooth halo, which may indicate a smaller contrast in progenitor mass and chemical enrichment between the inner and outer halo than in the Milky Way. Alternatively, relatively massive and chemically enriched progenitors may have contributed more strongly, with their debris extending to larger radii in the M31 smooth halo \citep[e.g.,][]{DSouza2018b, Gilbert2020}. In contrast, the M31 smooth halo shows a stronger positive [$\alpha$/Fe] gradient than the Milky Way. More accurate [$\alpha$/Fe] estimates for M31 and measurements of Mg and other $\alpha$-elements farther into the Milky Way halo are needed to better compare the progenitor SFHs of the two galaxies. The ongoing `{\=O}nohi`ula Subaru PFS spectroscopic survey and future HSC $NB395$ photometry will enable more precise abundance measurements for a larger M31 halo sample and provide tighter constraints on its star formation and assembly history.

\begin{acknowledgements}
The authors express thanks to Alex Dowling, Ke Wang, Christopher Sweet, Charles Vardeman, Karoline Gilbert, and Ivanna Escala for their input and for sharing data.

This research is based in part on data collected at the Subaru Telescope, which is operated by the National Astronomical Observatory of Japan. 
This work was supported in part by the JSPS KAKENHI Grant (No. JP24K00669, JP25K24687) for M.C. 
T.C.B. acknowledges partial support from grants PHY 14-30152; Physics Frontier Center/JINA Center for the Evolution of the Elements (JINA-CEE), and OISE-1927130; The International Research Network for Nuclear Astrophysics (IReNA), awarded by the US National Science Foundation, and DE-SC0023128; the Center for Nuclear Astrophysics Across Messengers (CeNAM), awarded by the U.S. Department of Energy, Office of Science, Office of Nuclear Physics. Y.S.L. acknowledges support from the National Research Foundation (NRF) of Korea grant funded by the Ministry of Science and ICT (RS-2024-00333766).  J.H., E.N.K., and M.C.\ gratefully acknowledge support from Notre Dame Global Asia Research Collaboration Grants.

Portions of this text were written with the assistance of ChatGPT \citep{openai}, especially regarding English usage and grammar.

\end{acknowledgements}

\begin{contribution}
J.~Hong led the scientific analysis and manuscript writing.
E.~Kirby advised on the scientific content.
R.~Gerasimov contributed to the development of Chemfit and helped with its use in generating synthetic data.
M.~Chiba, Y.~Komiyama, and I.~Ogami provided the data and suggestions for improving the manuscript.
T.~Tang provided guidance on the random forest analysis.
T.~Beers and Y.~S.~Lee provided feedback on the manuscript.
\end{contribution}

\vfill\eject

\gdef\thefigure{\thesection.\arabic{figure}}    

\bibliography{main}{}

\begin{thebibliography}{}
\expandafter\ifx\csname natexlab\endcsname\relax\def\natexlab#1{#1}\fi
\providecommand{\url}[1]{\href{#1}{#1}}
\providecommand{\dodoi}[1]{doi:~\href{http://doi.org/#1}{\nolinkurl{#1}}}
\providecommand{\doeprint}[1]{\href{http://ascl.net/#1}{\nolinkurl{http://ascl.net/#1}}}
\providecommand{\doarXiv}[1]{\href{https://arxiv.org/abs/#1}{\nolinkurl{https://arxiv.org/abs/#1}}}

\bibitem[{C.~P. {Ahn} {et~al.}(2014){Ahn}, {Alexandroff}, {Allende Prieto}, {Anders}, {Anderson}, {Anderton}, {Andrews}, {Aubourg}, {Bailey}, {Bastien}, {Bautista}, {Beers}, {Beifiori}, {Bender}, {Berlind}, {Beutler}, {Bhardwaj}, {Bird}, {Bizyaev}, {Blake}, {Blanton}, {Blomqvist}, {Bochanski}, {Bolton}, {Borde}, {Bovy}, {Shelden Bradley}, {Brandt}, {Brauer}, {Brinkmann}, {Brownstein}, {Busca}, {Carithers}, {Carlberg}, {Carnero}, {Carr}, {Chiappini}, {Chojnowski}, {Chuang}, {Comparat}, {Crepp}, {Cristiani}, {Croft}, {Cuesta}, {Cunha}, {da Costa}, {Dawson}, {De Lee}, {Dean}, {Delubac}, {Deshpande}, {Dhital}, {Ealet}, {Ebelke}, {Edmondson}, {Eisenstein}, {Epstein}, {Escoffier}, {Esposito}, {Evans}, {Fabbian}, {Fan}, {Favole}, {Femen{\'\i}a Castell{\'a}}, {Fern{\'a}ndez Alvar}, {Feuillet}, {Filiz Ak}, {Finley}, {Fleming}, {Font-Ribera}, {Frinchaboy}, {Galbraith-Frew}, {Garc{\'\i}a-Hern{\'a}ndez}, {Garc{\'\i}a P{\'e}rez}, {Ge}, {G{\'e}nova-Santos}, {Gillespie}, {Girardi}, {Gonz{\'a}lez Hern{\'a}ndez}, {Gott},
  {Gunn}, {Guo}, {Halverson}, {Harding}, {Harris}, {Hasselquist}, {Hawley}, {Hayden}, {Hearty}, {Herrero Dav{\'o}}, {Ho}, {Hogg}, {Holtzman}, {Honscheid}, {Huehnerhoff}, {Ivans}, {Jackson}, {Jiang}, {Johnson}, {Kinemuchi}, {Kirkby}, {Klaene}, {Kneib}, {Koesterke}, {Lan}, {Lang}, {Le Goff}, {Leauthaud}, {Lee}, {Lee}, {Long}, {Loomis}, {Lucatello}, {Lupton}, {Ma}, {Mack}, {Mahadevan}, {Maia}, {Majewski}, {Malanushenko}, {Malanushenko}, {Manchado}, {Manera}, {Maraston}, {Margala}, {Martell}, {Masters}, {McBride}, {McGreer}, {McMahon}, {M{\'e}nard}, {M{\'e}sz{\'a}ros}, {Miralda-Escud{\'e}}, {Miyatake}, {Montero-Dorta}, {Montesano}, {More}, {Morrison}, {Muna}, {Munn}, {Myers}, {Nguyen}, {Nichol}, {Nidever}, {Noterdaeme}, {Nuza}, {O'Connell}, {O'Connell}, {O'Connell}, {Olmstead}, {Oravetz}, {Owen}, {Padmanabhan}, {Palanque-Delabrouille}, {Pan}, {Parejko}, {Parihar}, {P{\^a}ris}, {Pepper}, {Percival}, {P{\'e}rez-R{\`a}fols}, {Dotto Perottoni}, {Petitjean}, {Pieri}, {Pinsonneault}, {Prada}, {Price-Whelan}, {Raddick},
  {Rahman}, {Rebolo}, {Reid}, {Richards}, {Riffel}, {Robin}, {Rocha-Pinto}, {Rockosi}, {Roe}, {Ross}, {Ross}, {Rossi}, {Roy}, {Rubi{\~n}o-Martin}, {Sabiu}, {S{\'a}nchez}, {Santiago}, {Sayres}, {Schiavon}, {Schlegel}, {Schlesinger}, {Schmidt}, {Schneider}, {Schultheis}, {Sellgren}, {Seo}, {Shen}, {Shetrone}, {Shu}, {Simmons}, {Skrutskie}, \& {Slosar}}]{Ahn2014}
{Ahn}, C.~P., {Alexandroff}, R., {Allende Prieto}, C., {et~al.} 2014, \bibinfo{title}{{The Tenth Data Release of the Sloan Digital Sky Survey: First Spectroscopic Data from the SDSS-III Apache Point Observatory Galactic Evolution Experiment},} \apjs, 211, 17, \dodoi{10.1088/0067-0049/211/2/17}

\bibitem[{S. {Bhattacharya} {et~al.}(2019){Bhattacharya}, {Arnaboldi}, {Caldwell}, {Gerhard}, {Bla{\~n}a}, {McConnachie}, {Hartke}, {Guhathakurta}, {Pulsoni}, \& {Freeman}}]{Bhattacharya2019}
{Bhattacharya}, S., {Arnaboldi}, M., {Caldwell}, N., {et~al.} 2019, \bibinfo{title}{{The survey of planetary nebulae in Andromeda (M 31). II. Age-velocity dispersion relation in the disc from planetary nebulae},} \aap, 631, A56, \dodoi{10.1051/0004-6361/201935898}

\bibitem[{S. {Bhattacharya} {et~al.}(2022){Bhattacharya}, {Arnaboldi}, {Caldwell}, {Gerhard}, {Kobayashi}, {Hartke}, {Freeman}, {McConnachie}, \& {Guhathakurta}}]{Bhattacharya2022}
{Bhattacharya}, S., {Arnaboldi}, M., {Caldwell}, N., {et~al.} 2022, \bibinfo{title}{{The survey of planetary nebulae in Andromeda (M31) - IV. Radial oxygen and argon abundance gradients of the thin and thicker disc},} \mnras, 517, 2343, \dodoi{10.1093/mnras/stac2703}

\bibitem[{G. {Biau} \& E. {Scornet}(2015){Biau} \& {Scornet}}]{Biau2015}
{Biau}, G., \& {Scornet}, E. 2015, \bibinfo{title}{{A Random Forest Guided Tour},} arXiv e-prints, arXiv:1511.05741, \dodoi{10.48550/arXiv.1511.05741}

\bibitem[{C.~M. {Brasseur} {et~al.}(2011){Brasseur}, {Martin}, {Rix}, {Irwin}, {Ferguson}, {McConnachie}, \& {de Jong}}]{Brasseur2011}
{Brasseur}, C.~M., {Martin}, N.~F., {Rix}, H.-W., {et~al.} 2011, \bibinfo{title}{{A Deep Photometric Look at Two of Andromeda's Dwarf Spheroidals: X and XVII},} \apj, 729, 23, \dodoi{10.1088/0004-637X/729/1/23}

\bibitem[{L. {Breiman}(2001){Breiman}}]{Breiman2001}
{Breiman}, L. 2001, \bibinfo{title}{{Random Forests.},} Machine Learning, 45, 5, \dodoi{10.1023/A:1010933404324}

\bibitem[{J.~S. {Bullock} \& K.~V. {Johnston}(2005){Bullock} \& {Johnston}}]{Bullock2005}
{Bullock}, J.~S., \& {Johnston}, K.~V. 2005, \bibinfo{title}{{Tracing Galaxy Formation with Stellar Halos. I. Methods},} \apj, 635, 931, \dodoi{10.1086/497422}

\bibitem[{F. {Castelli} \& R.~L. {Kurucz}(2003){Castelli} \& {Kurucz}}]{Castelli2003}
{Castelli}, F., \& {Kurucz}, R.~L. 2003, \bibinfo{title}{{New Grids of ATLAS9 Model Atmospheres},} in IAU Symposium, Vol. 210, Modelling of Stellar Atmospheres, ed. N.~{Piskunov}, W.~W. {Weiss}, \& D.~F. {Gray}, A20, \dodoi{10.48550/arXiv.astro-ph/0405087}

\bibitem[{S.~C. {Chapman} {et~al.}(2006){Chapman}, {Ibata}, {Lewis}, {Ferguson}, {Irwin}, {McConnachie}, \& {Tanvir}}]{Chapman2006}
{Chapman}, S.~C., {Ibata}, R., {Lewis}, G.~F., {et~al.} 2006, \bibinfo{title}{{A Kinematically Selected, Metal-poor Stellar Halo in the Outskirts of M31},} \apj, 653, 255, \dodoi{10.1086/508599}

\bibitem[{J. {Choi} {et~al.}(2016){Choi}, {Dotter}, {Conroy}, {Cantiello}, {Paxton}, \& {Johnson}}]{Choi2016}
{Choi}, J., {Dotter}, A., {Conroy}, C., {et~al.} 2016, \bibinfo{title}{{Mesa Isochrones and Stellar Tracks (MIST). I. Solar-scaled Models},} \apj, 823, 102, \dodoi{10.3847/0004-637X/823/2/102}

\bibitem[{M.~L.~M. {Collins} {et~al.}(2010){Collins}, {Chapman}, {Irwin}, {Martin}, {Ibata}, {Zucker}, {Blain}, {Ferguson}, {Lewis}, {McConnachie}, \& {Pe{\~n}arrubia}}]{Collins2010}
{Collins}, M.~L.~M., {Chapman}, S.~C., {Irwin}, M.~J., {et~al.} 2010, \bibinfo{title}{{A Keck/DEIMOS spectroscopic survey of the faint M31 satellites AndIX, AndXI, AndXII and AndXIII{\textdagger}},} \mnras, 407, 2411, \dodoi{10.1111/j.1365-2966.2010.17069.x}

\bibitem[{M.~L.~M. {Collins} {et~al.}(2011){Collins}, {Chapman}, {Rich}, {Irwin}, {Pe{\~n}arrubia}, {Ibata}, {Arimoto}, {Brooks}, {Ferguson}, {Lewis}, {McConnachie}, \& {Venn}}]{Collins2011}
{Collins}, M.~L.~M., {Chapman}, S.~C., {Rich}, R.~M., {et~al.} 2011, \bibinfo{title}{{The scatter about the 'Universal' dwarf spheroidal mass profile: a kinematic study of the M31 satellites And V and And VI},} \mnras, 417, 1170, \dodoi{10.1111/j.1365-2966.2011.19342.x}

\bibitem[{S. {Courteau} {et~al.}(2011){Courteau}, {Widrow}, {McDonald}, {Guhathakurta}, {Gilbert}, {Zhu}, {Beaton}, \& {Majewski}}]{Courteau2011}
{Courteau}, S., {Widrow}, L.~M., {McDonald}, M., {et~al.} 2011, \bibinfo{title}{{The Luminosity Profile and Structural Parameters of the Andromeda Galaxy},} \apj, 739, 20, \dodoi{10.1088/0004-637X/739/1/20}

\bibitem[{J.~J. {Dalcanton} {et~al.}(2012){Dalcanton}, {Williams}, {Lang}, {Lauer}, {Kalirai}, {Seth}, {Dolphin}, {Rosenfield}, {Weisz}, {Bell}, {Bianchi}, {Boyer}, {Caldwell}, {Dong}, {Dorman}, {Gilbert}, {Girardi}, {Gogarten}, {Gordon}, {Guhathakurta}, {Hodge}, {Holtzman}, {Johnson}, {Larsen}, {Lewis}, {Melbourne}, {Olsen}, {Rix}, {Rosema}, {Saha}, {Sarajedini}, {Skillman}, \& {Stanek}}]{Dalcanton2012}
{Dalcanton}, J.~J., {Williams}, B.~F., {Lang}, D., {et~al.} 2012, \bibinfo{title}{{The Panchromatic Hubble Andromeda Treasury},} \apjs, 200, 18, \dodoi{10.1088/0067-0049/200/2/18}

\bibitem[{J.~J. {Dalcanton} {et~al.}(2023){Dalcanton}, {Bell}, {Choi}, {Dolphin}, {Fouesneau}, {Girardi}, {Hogg}, {Seth}, \& {Williams}}]{Dalcanton2023}
{Dalcanton}, J.~J., {Bell}, E.~F., {Choi}, Y., {et~al.} 2023, \bibinfo{title}{{The Panchromatic Hubble Andromeda Treasury. XX. The Disk of M31 is Thick},} \aj, 166, 80, \dodoi{10.3847/1538-3881/accc83}

\bibitem[{G. {de Vaucouleurs} {et~al.}(1991){de Vaucouleurs}, {de Vaucouleurs}, {Corwin}, {Buta}, {Paturel}, \& {Fouque}}]{deVaucouleurs1991}
{de Vaucouleurs}, G., {de Vaucouleurs}, A., {Corwin}, Jr., H.~G., {et~al.} 1991, {Third Reference Catalogue of Bright Galaxies}, Vol.~1

\bibitem[{K. {Ding} {et~al.}(2025){Ding}, {Filion}, {Wyse}, {Kirby}, {Ogami}, {Chiba}, {Komiyama}, {Dobos}, \& {Szalay}}]{Ding2025}
{Ding}, K., {Filion}, C., {Wyse}, R. F.~G., {et~al.} 2025, \bibinfo{title}{{Machine Learning Techniques to Distinguish Giant Stars from Dwarf Stars Using Only Photometry{\textemdash}Pushing Redward},} \aj, 170, 327, \dodoi{10.3847/1538-3881/ae10a7}

\bibitem[{A. {Dotter} {et~al.}(2008){Dotter}, {Chaboyer}, {Jevremovi{\'c}}, {Kostov}, {Baron}, \& {Ferguson}}]{Dotter2008}
{Dotter}, A., {Chaboyer}, B., {Jevremovi{\'c}}, D., {et~al.} 2008, \bibinfo{title}{{The Dartmouth Stellar Evolution Database},} \apjs, 178, 89, \dodoi{10.1086/589654}

\bibitem[{R. {D'Souza} \& E.~F. {Bell}(2018){D'Souza} \& {Bell}}]{DSouza2018b}
{D'Souza}, R., \& {Bell}, E.~F. 2018, \bibinfo{title}{{The masses and metallicities of stellar haloes reflect galactic merger histories},} \mnras, 474, 5300, \dodoi{10.1093/mnras/stx3081}

\bibitem[{I. {Escala} {et~al.}(2020{\natexlab{a}}){Escala}, {Gilbert}, {Kirby}, {Wojno}, {Cunningham}, \& {Guhathakurta}}]{Escala2020a}
{Escala}, I., {Gilbert}, K.~M., {Kirby}, E.~N., {et~al.} 2020{\natexlab{a}}, \bibinfo{title}{{Elemental Abundances in M31: A Comparative Analysis of Alpha and Iron Element Abundances in the the Outer Disk, Giant Stellar Stream, and Inner Halo of M31},} \apj, 889, 177, \dodoi{10.3847/1538-4357/ab6659}

\bibitem[{I. {Escala} {et~al.}(2019){Escala}, {Kirby}, {Gilbert}, {Cunningham}, \& {Wojno}}]{Escala2019}
{Escala}, I., {Kirby}, E.~N., {Gilbert}, K.~M., {Cunningham}, E.~C., \& {Wojno}, J. 2019, \bibinfo{title}{{Elemental Abundances in M31: Alpha and Iron Element Abundances from Low-resolution Resolved Stellar Spectroscopy in the Stellar Halo},} \apj, 878, 42, \dodoi{10.3847/1538-4357/ab1eac}

\bibitem[{I. {Escala} {et~al.}(2020{\natexlab{b}}){Escala}, {Kirby}, {Gilbert}, {Wojno}, {Cunningham}, \& {Guhathakurta}}]{Escala2020b}
{Escala}, I., {Kirby}, E.~N., {Gilbert}, K.~M., {et~al.} 2020{\natexlab{b}}, \bibinfo{title}{{Elemental Abundances in M31: Properties of the Inner Stellar Halo},} \apj, 902, 51, \dodoi{10.3847/1538-4357/abb474}

\bibitem[{M.~A. {Fardal} {et~al.}(2007){Fardal}, {Guhathakurta}, {Babul}, \& {McConnachie}}]{Fardal2007}
{Fardal}, M.~A., {Guhathakurta}, P., {Babul}, A., \& {McConnachie}, A.~W. 2007, \bibinfo{title}{{Investigating the Andromeda stream - III. A young shell system in M31},} \mnras, 380, 15, \dodoi{10.1111/j.1365-2966.2007.11929.x}

\bibitem[{A.~M.~N. {Ferguson} {et~al.}(2002){Ferguson}, {Irwin}, {Ibata}, {Lewis}, \& {Tanvir}}]{Ferguson2002}
{Ferguson}, A. M.~N., {Irwin}, M.~J., {Ibata}, R.~A., {Lewis}, G.~F., \& {Tanvir}, N.~R. 2002, \bibinfo{title}{{Evidence for Stellar Substructure in the Halo and Outer Disk of M31},} \aj, 124, 1452, \dodoi{10.1086/342019}

\bibitem[{E. {Fern{\'a}ndez-Alvar} {et~al.}(2015){Fern{\'a}ndez-Alvar}, {Allende Prieto}, {Schlesinger}, {Beers}, {Robin}, {Schneider}, {Lee}, {Bizyaev}, {Ebelke}, {Malanushenko}, {Malanushenko}, {Oravetz}, {Pan}, \& {Simmons}}]{Fernandez-Alvar2015}
{Fern{\'a}ndez-Alvar}, E., {Allende Prieto}, C., {Schlesinger}, K.~J., {et~al.} 2015, \bibinfo{title}{{Deep SDSS optical spectroscopy of distant halo stars. II. Iron, calcium, and magnesium abundances},} \aap, 577, A81, \dodoi{10.1051/0004-6361/201425455}

\bibitem[{H. {Furusawa} {et~al.}(2018){Furusawa}, {Koike}, {Takata}, {Okura}, {Miyatake}, {Lupton}, {Bickerton}, {Price}, {Bosch}, {Yasuda}, {Mineo}, {Yamada}, {Miyazaki}, {Nakata}, {Koshida}, {Komiyama}, {Utsumi}, {Kawanomoto}, {Jeschke}, {Noumaru}, {Schubert}, {Iwata}, {Finet}, {Fujiyoshi}, {Tajitsu}, {Terai}, \& {Lee}}]{Furusawa2018}
{Furusawa}, H., {Koike}, M., {Takata}, T., {et~al.} 2018, \bibinfo{title}{{The on-site quality-assurance system for Hyper Suprime-Cam: OSQAH},} \pasj, 70, S3, \dodoi{10.1093/pasj/psx079}

\bibitem[{R. {Gerasimov} {et~al.}(2022){Gerasimov}, {Burgasser}, {Homeier}, {Bedin}, {Rees}, {Scalco}, {Anderson}, \& {Salaris}}]{Gerasimov2022}
{Gerasimov}, R., {Burgasser}, A.~J., {Homeier}, D., {et~al.} 2022, \bibinfo{title}{{The HST Large Program on {\ensuremath{\omega}} Centauri. V. Exploring the Ultracool Dwarf Population with Stellar Atmosphere and Evolutionary Modeling},} \apj, 930, 24, \dodoi{10.3847/1538-4357/ac61e5}

\bibitem[{K.~M. {Gilbert} {et~al.}(2020){Gilbert}, {Wojno}, {Kirby}, {Escala}, {Beaton}, {Guhathakurta}, \& {Majewski}}]{Gilbert2020}
{Gilbert}, K.~M., {Wojno}, J., {Kirby}, E.~N., {et~al.} 2020, \bibinfo{title}{{Elemental Abundances in M31: Iron and Alpha Element Abundances in M31's Outer Halo},} \aj, 160, 41, \dodoi{10.3847/1538-3881/ab9602}

\bibitem[{K.~M. {Gilbert} {et~al.}(2007){Gilbert}, {Fardal}, {Kalirai}, {Guhathakurta}, {Geha}, {Isler}, {Majewski}, {Ostheimer}, {Patterson}, {Reitzel}, {Kirby}, \& {Cooper}}]{Gilbert2007}
{Gilbert}, K.~M., {Fardal}, M., {Kalirai}, J.~S., {et~al.} 2007, \bibinfo{title}{{Stellar Kinematics in the Complicated Inner Spheroid of M31: Discovery of Substructure along the Southeastern Minor Axis and Its Relationship to the Giant Southern Stream},} \apj, 668, 245, \dodoi{10.1086/521094}

\bibitem[{K.~M. {Gilbert} {et~al.}(2012){Gilbert}, {Guhathakurta}, {Beaton}, {Bullock}, {Geha}, {Kalirai}, {Kirby}, {Majewski}, {Ostheimer}, {Patterson}, {Tollerud}, {Tanaka}, \& {Chiba}}]{Gilbert2012}
{Gilbert}, K.~M., {Guhathakurta}, P., {Beaton}, R.~L., {et~al.} 2012, \bibinfo{title}{{Global Properties of M31's Stellar Halo from the SPLASH Survey. I. Surface Brightness Profile},} \apj, 760, 76, \dodoi{10.1088/0004-637X/760/1/76}

\bibitem[{K.~M. {Gilbert} {et~al.}(2014){Gilbert}, {Kalirai}, {Guhathakurta}, {Beaton}, {Geha}, {Kirby}, {Majewski}, {Patterson}, {Tollerud}, {Bullock}, {Tanaka}, \& {Chiba}}]{Gilbert2014}
{Gilbert}, K.~M., {Kalirai}, J.~S., {Guhathakurta}, P., {et~al.} 2014, \bibinfo{title}{{Global Properties of M31's Stellar Halo from the SPLASH Survey. II. Metallicity Profile},} \apj, 796, 76, \dodoi{10.1088/0004-637X/796/2/76}

\bibitem[{T.~J. {Hastie} {et~al.}(2009){Hastie}, {Tibshirani}, \& {Friedman}}]{Hastie2009}
{Hastie}, T.~J., {Tibshirani}, R.~J., \& {Friedman}, J.~H. 2009, \bibinfo{title}{{The Elements of Statistical Learning: Data Mining, Inference, and Prediction},} Springer

\bibitem[{J. {Hong} {et~al.}(2025){Hong}, {Kirby}, {Tang}, {Chiba}, {Komiyama}, {Henderson}, {Ogami}, \& {Beers}}]{Hong2025}
{Hong}, J., {Kirby}, E.~N., {Tang}, T.~M., {et~al.} 2025, \bibinfo{title}{{Estimating Fe and Mg Abundances in the Milky Way Dwarf Galaxies Using Subaru/HSC and DEIMOS},} \apj, 989, 48, \dodoi{10.3847/1538-4357/ade679}

\bibitem[{R. {Ibata} {et~al.}(2001){Ibata}, {Irwin}, {Lewis}, {Ferguson}, \& {Tanvir}}]{Ibata2001}
{Ibata}, R., {Irwin}, M., {Lewis}, G., {Ferguson}, A. M.~N., \& {Tanvir}, N. 2001, \bibinfo{title}{{A giant stream of metal-rich stars in the halo of the galaxy M31},} \nat, 412, 49, \dodoi{10.1038/35083506}

\bibitem[{R. {Ibata} {et~al.}(2007){Ibata}, {Martin}, {Irwin}, {Chapman}, {Ferguson}, {Lewis}, \& {McConnachie}}]{Ibata2007}
{Ibata}, R., {Martin}, N.~F., {Irwin}, M., {et~al.} 2007, \bibinfo{title}{{The Haunted Halos of Andromeda and Triangulum: A Panorama of Galaxy Formation in Action},} \apj, 671, 1591, \dodoi{10.1086/522574}

\bibitem[{R.~A. {Ibata} {et~al.}(2014){Ibata}, {Lewis}, {McConnachie}, {Martin}, {Irwin}, {Ferguson}, {Babul}, {Bernard}, {Chapman}, {Collins}, {Fardal}, {Mackey}, {Navarro}, {Pe{\~n}arrubia}, {Rich}, {Tanvir}, \& {Widrow}}]{Ibata2014}
{Ibata}, R.~A., {Lewis}, G.~F., {McConnachie}, A.~W., {et~al.} 2014, \bibinfo{title}{{The Large-scale Structure of the Halo of the Andromeda Galaxy. I. Global Stellar Density, Morphology and Metallicity Properties},} \apj, 780, 128, \dodoi{10.1088/0004-637X/780/2/128}

\bibitem[{J.~S. {Kalirai} {et~al.}(2006){Kalirai}, {Gilbert}, {Guhathakurta}, {Majewski}, {Ostheimer}, {Rich}, {Cooper}, {Reitzel}, \& {Patterson}}]{Kalirai2006}
{Kalirai}, J.~S., {Gilbert}, K.~M., {Guhathakurta}, P., {et~al.} 2006, \bibinfo{title}{{The Metal-poor Halo of the Andromeda Spiral Galaxy (M31)1,},} \apj, 648, 389, \dodoi{10.1086/505697}

\bibitem[{Y. {Kang} {et~al.}(2009){Kang}, {Bianchi}, \& {Rey}}]{Kang2009}
{Kang}, Y., {Bianchi}, L., \& {Rey}, S.-C. 2009, \bibinfo{title}{{An Ultraviolet Study of Star-Forming Regions in M31},} \apj, 703, 614, \dodoi{10.1088/0004-637X/703/1/614}

\bibitem[{S. {Kawanomoto} {et~al.}(2018){Kawanomoto}, {Uraguchi}, {Komiyama}, {Miyazaki}, {Furusawa}, {Finet}, {Hattori}, {Wang}, {Yasuda}, \& {Suzuki}}]{Kawanomoto2018}
{Kawanomoto}, S., {Uraguchi}, F., {Komiyama}, Y., {et~al.} 2018, \bibinfo{title}{{Hyper Suprime-Cam: Filters},} \pasj, 70, 66, \dodoi{10.1093/pasj/psy056}

\bibitem[{E.~N. {Kirby} {et~al.}(2013){Kirby}, {Cohen}, {Guhathakurta}, {Cheng}, {Bullock}, \& {Gallazzi}}]{Kirby2013}
{Kirby}, E.~N., {Cohen}, J.~G., {Guhathakurta}, P., {et~al.} 2013, \bibinfo{title}{{The Universal Stellar Mass-Stellar Metallicity Relation for Dwarf Galaxies},} \apj, 779, 102, \dodoi{10.1088/0004-637X/779/2/102}

\bibitem[{E.~N. {Kirby} {et~al.}(2020){Kirby}, {Gilbert}, {Escala}, {Wojno}, {Guhathakurta}, {Majewski}, \& {Beaton}}]{Kirby2020}
{Kirby}, E.~N., {Gilbert}, K.~M., {Escala}, I., {et~al.} 2020, \bibinfo{title}{{Elemental Abundances in M31: The Kinematics and Chemical Evolution of Dwarf Spheroidal Satellite Galaxies},} \aj, 159, 46, \dodoi{10.3847/1538-3881/ab5f0f}

\bibitem[{C. {Kobayashi} {et~al.}(2023){Kobayashi}, {Bhattacharya}, {Arnaboldi}, \& {Gerhard}}]{Kobayashi2023}
{Kobayashi}, C., {Bhattacharya}, S., {Arnaboldi}, M., \& {Gerhard}, O. 2023, \bibinfo{title}{{On the {\ensuremath{\alpha}}/Fe Bimodality of the M31 Disks},} \apjl, 956, L14, \dodoi{10.3847/2041-8213/acf7c7}

\bibitem[{A. {Koch} {et~al.}(2008){Koch}, {Rich}, {Reitzel}, {Martin}, {Ibata}, {Chapman}, {Majewski}, {Mori}, {Loh}, {Ostheimer}, \& {Tanaka}}]{Koch2008}
{Koch}, A., {Rich}, R.~M., {Reitzel}, D.~B., {et~al.} 2008, \bibinfo{title}{{Kinematic and Chemical Constraints on the Formation of M31's Inner and Outer Halo},} \apj, 689, 958, \dodoi{10.1086/592373}

\bibitem[{Y. {Komiyama} {et~al.}(2018{\natexlab{a}}){Komiyama}, {Obuchi}, {Nakaya}, {Kamata}, {Kawanomoto}, {Utsumi}, {Miyazaki}, {Uraguchi}, {Furusawa}, {Morokuma}, {Uchida}, {Miyatake}, {Mineo}, {Fujimori}, {Aihara}, {Karoji}, {Gunn}, \& {Wang}}]{Komiyama2018b}
{Komiyama}, Y., {Obuchi}, Y., {Nakaya}, H., {et~al.} 2018{\natexlab{a}}, \bibinfo{title}{{Hyper Suprime-Cam: Camera dewar design},} \pasj, 70, S2, \dodoi{10.1093/pasj/psx069}

\bibitem[{Y. {Komiyama} {et~al.}(2018{\natexlab{b}}){Komiyama}, {Chiba}, {Tanaka}, {Tanaka}, {Kirihara}, {Miki}, {Mori}, {Lupton}, {Guhathakurta}, {Kalirai}, {Gilbert}, {Kirby}, {Lee}, {Jang}, {Sharma}, \& {Hayashi}}]{Komiyama2018a}
{Komiyama}, Y., {Chiba}, M., {Tanaka}, M., {et~al.} 2018{\natexlab{b}}, \bibinfo{title}{{Stellar Stream and Halo Structure in the Andromeda Galaxy from a Subaru/Hyper Suprime-Cam Survey},} \apj, 853, 29, \dodoi{10.3847/1538-4357/aaa129}

\bibitem[{B. Lu \& J. Hardin(2021)Lu \& Hardin}]{Lu2021}
Lu, B., \& Hardin, J. 2021, \bibinfo{title}{A unified framework for random forest prediction error estimi ation,} Journal of Machine Learning Research, 22, 1

\bibitem[{S.~R. {Majewski} {et~al.}(2000){Majewski}, {Ostheimer}, {Kunkel}, \& {Patterson}}]{Majewski2000}
{Majewski}, S.~R., {Ostheimer}, J.~C., {Kunkel}, W.~E., \& {Patterson}, R.~J. 2000, \bibinfo{title}{{Exploring Halo Substructure with Giant Stars. I. Survey Description and Calibration of the Photometric Search Technique},} \aj, 120, 2550, \dodoi{10.1086/316836}

\bibitem[{A.~W. {McConnachie}(2012){McConnachie}}]{McConnachie2012}
{McConnachie}, A.~W. 2012, \bibinfo{title}{{The Observed Properties of Dwarf Galaxies in and around the Local Group},} \aj, 144, 4, \dodoi{10.1088/0004-6256/144/1/4}

\bibitem[{A.~W. {McConnachie} \& M.~J. {Irwin}(2006){McConnachie} \& {Irwin}}]{McConnachie2006}
{McConnachie}, A.~W., \& {Irwin}, M.~J. 2006, \bibinfo{title}{{Structural properties of the M31 dwarf spheroidal galaxies},} \mnras, 365, 1263, \dodoi{10.1111/j.1365-2966.2005.09806.x}

\bibitem[{A.~W. {McConnachie} {et~al.}(2005){McConnachie}, {Irwin}, {Ferguson}, {Ibata}, {Lewis}, \& {Tanvir}}]{McConnachie2005}
{McConnachie}, A.~W., {Irwin}, M.~J., {Ferguson}, A.~M.~N., {et~al.} 2005, \bibinfo{title}{{Distances and metallicities for 17 Local Group galaxies},} \mnras, 356, 979, \dodoi{10.1111/j.1365-2966.2004.08514.x}

\bibitem[{A.~W. {McConnachie} {et~al.}(2009){McConnachie}, {Irwin}, {Ibata}, {Dubinski}, {Widrow}, {Martin}, {C{\^o}t{\'e}}, {Dotter}, {Navarro}, {Ferguson}, {Puzia}, {Lewis}, {Babul}, {Barmby}, {Bienaym{\'e}}, {Chapman}, {Cockcroft}, {Collins}, {Fardal}, {Harris}, {Huxor}, {Mackey}, {Pe{\~n}arrubia}, {Rich}, {Richer}, {Siebert}, {Tanvir}, {Valls-Gabaud}, \& {Venn}}]{McConnachie2009}
{McConnachie}, A.~W., {Irwin}, M.~J., {Ibata}, R.~A., {et~al.} 2009, \bibinfo{title}{{The remnants of galaxy formation from a panoramic survey of the region around M31},} \nat, 461, 66, \dodoi{10.1038/nature08327}

\bibitem[{A.~W. {McConnachie} {et~al.}(2018){McConnachie}, {Ibata}, {Martin}, {Ferguson}, {Collins}, {Gwyn}, {Irwin}, {Lewis}, {Mackey}, {Davidge}, {Arias}, {Conn}, {C{\^o}t{\'e}}, {Crnojevic}, {Huxor}, {Penarrubia}, {Spengler}, {Tanvir}, {Valls-Gabaud}, {Babul}, {Barmby}, {Bate}, {Bernard}, {Chapman}, {Dotter}, {Harris}, {McMonigal}, {Navarro}, {Puzia}, {Rich}, {Thomas}, \& {Widrow}}]{McConnachie2018}
{McConnachie}, A.~W., {Ibata}, R., {Martin}, N., {et~al.} 2018, \bibinfo{title}{{The Large-scale Structure of the Halo of the Andromeda Galaxy. II. Hierarchical Structure in the Pan-Andromeda Archaeological Survey},} \apj, 868, 55, \dodoi{10.3847/1538-4357/aae8e7}

\bibitem[{S. {Miyazaki} {et~al.}(2012){Miyazaki}, {Komiyama}, {Nakaya}, {Kamata}, {Doi}, {Hamana}, {Karoji}, {Furusawa}, {Kawanomoto}, {Morokuma}, {Ishizuka}, {Nariai}, {Tanaka}, {Uraguchi}, {Utsumi}, {Obuchi}, {Okura}, {Oguri}, {Takata}, {Tomono}, {Kurakami}, {Namikawa}, {Usuda}, {Yamanoi}, {Terai}, {Uekiyo}, {Yamada}, {Koike}, {Aihara}, {Fujimori}, {Mineo}, {Miyatake}, {Yasuda}, {Nishizawa}, {Saito}, {Tanaka}, {Uchida}, {Katayama}, {Wang}, {Chen}, {Lupton}, {Loomis}, {Bickerton}, {Price}, {Gunn}, {Suzuki}, {Miyazaki}, {Muramatsu}, {Yamamoto}, {Endo}, {Ezaki}, {Itoh}, {Miwa}, {Yokota}, {Matsuda}, {Ebinuma}, \& {Takeshi}}]{Miyazaki2012}
{Miyazaki}, S., {Komiyama}, Y., {Nakaya}, H., {et~al.} 2012, \bibinfo{title}{{Hyper Suprime-Cam},} in Society of Photo-Optical Instrumentation Engineers (SPIE) Conference Series, Vol. 8446, Ground-based and Airborne Instrumentation for Astronomy IV, ed. I.~S. {McLean}, S.~K. {Ramsay}, \& H.~{Takami}, 84460Z, \dodoi{10.1117/12.926844}

\bibitem[{S. {Miyazaki} {et~al.}(2018){Miyazaki}, {Komiyama}, {Kawanomoto}, {Doi}, {Furusawa}, {Hamana}, {Hayashi}, {Ikeda}, {Kamata}, {Karoji}, {Koike}, {Kurakami}, {Miyama}, {Morokuma}, {Nakata}, {Namikawa}, {Nakaya}, {Nariai}, {Obuchi}, {Oishi}, {Okada}, {Okura}, {Tait}, {Takata}, {Tanaka}, {Tanaka}, {Terai}, {Tomono}, {Uraguchi}, {Usuda}, {Utsumi}, {Yamada}, {Yamanoi}, {Aihara}, {Fujimori}, {Mineo}, {Miyatake}, {Oguri}, {Uchida}, {Tanaka}, {Yasuda}, {Takada}, {Murayama}, {Nishizawa}, {Sugiyama}, {Chiba}, {Futamase}, {Wang}, {Chen}, {Ho}, {Liaw}, {Chiu}, {Ho}, {Lai}, {Lee}, {Jeng}, {Iwamura}, {Armstrong}, {Bickerton}, {Bosch}, {Gunn}, {Lupton}, {Loomis}, {Price}, {Smith}, {Strauss}, {Turner}, {Suzuki}, {Miyazaki}, {Muramatsu}, {Yamamoto}, {Endo}, {Ezaki}, {Ito}, {Kawaguchi}, {Sofuku}, {Taniike}, {Akutsu}, {Dojo}, {Kasumi}, {Matsuda}, {Imoto}, {Miwa}, {Suzuki}, {Takeshi}, \& {Yokota}}]{Miyazaki2018}
{Miyazaki}, S., {Komiyama}, Y., {Kawanomoto}, S., {et~al.} 2018, \bibinfo{title}{{Hyper Suprime-Cam: System design and verification of image quality},} \pasj, 70, S1, \dodoi{10.1093/pasj/psx063}

\bibitem[{I. {Ogami} {et~al.}(2025){Ogami}, {Tanaka}, {Komiyama}, {Chiba}, {Guhathakurta}, {Kirby}, {Wyse}, {Filion}, {Gilbert}, {Escala}, {Mori}, {Kirihara}, {Tanaka}, {Ishigaki}, {Hayashi}, {Lee}, {Sharma}, {Kalirai}, \& {Lupton}}]{Ogami2025}
{Ogami}, I., {Tanaka}, M., {Komiyama}, Y., {et~al.} 2025, \bibinfo{title}{{The structure of the stellar halo of the Andromeda galaxy explored with the NB515 for Subaru/HSC - I. New insights on the stellar halo up to 120 kpc},} \mnras, 536, 530, \dodoi{10.1093/mnras/stae2527}

\bibitem[{S. {Okamoto} {et~al.}(2024){Okamoto}, {Ferguson}, {Arimoto}, {Ogami}, {{\v{Z}}emaitis}, {Chiba}, {Irwin}, {Jang}, {Koda}, {Komiyama}, {Lee}, {Lee}, {Rich}, {Tanaka}, \& {Tanaka}}]{Okamoto2024}
{Okamoto}, S., {Ferguson}, A. M.~N., {Arimoto}, N., {et~al.} 2024, \bibinfo{title}{{Uncovering the Ghostly Remains of an Extremely Diffuse Satellite in the Remote Halo of NGC 253},} \apjl, 967, L24, \dodoi{10.3847/2041-8213/ad4358}

\bibitem[{{OpenAI,}(2024)}]{openai}
{OpenAI,}. 2024, \bibinfo{title}{{},} ChatGPT-4o (May 13 version) [Large Language Model], https://chat.openai.com/chat

\bibitem[{V.~M. {Placco} {et~al.}(2014){Placco}, {Frebel}, {Beers}, \& {Stancliffe}}]{Placco2014}
{Placco}, V.~M., {Frebel}, A., {Beers}, T.~C., \& {Stancliffe}, R.~J. 2014, \bibinfo{title}{{Carbon-enhanced Metal-poor Star Frequencies in the Galaxy: Corrections for the Effect of Evolutionary Status on Carbon Abundances},} \apj, 797, 21, \dodoi{10.1088/0004-637X/797/1/21}

\bibitem[{J.~C. {Richardson} {et~al.}(2011){Richardson}, {Irwin}, {McConnachie}, {Martin}, {Dotter}, {Ferguson}, {Ibata}, {Chapman}, {Lewis}, {Tanvir}, \& {Rich}}]{Richardson2011}
{Richardson}, J.~C., {Irwin}, M.~J., {McConnachie}, A.~W., {et~al.} 2011, \bibinfo{title}{{PAndAS' Progeny: Extending the M31 Dwarf Galaxy Cabal},} \apj, 732, 76, \dodoi{10.1088/0004-637X/732/2/76}

\bibitem[{M. {Tanaka} {et~al.}(2010){Tanaka}, {Chiba}, {Komiyama}, {Guhathakurta}, {Kalirai}, \& {Iye}}]{Tanaka2010}
{Tanaka}, M., {Chiba}, M., {Komiyama}, Y., {et~al.} 2010, \bibinfo{title}{{Structure and Population of the Andromeda Stellar Halo from a Subaru/Suprime-Cam Survey},} \apj, 708, 1168, \dodoi{10.1088/0004-637X/708/2/1168}

\bibitem[{D.~A. {VandenBerg} {et~al.}(2006){VandenBerg}, {Bergbusch}, \& {Dowler}}]{VandenBerg2006}
{VandenBerg}, D.~A., {Bergbusch}, P.~A., \& {Dowler}, P.~D. 2006, \bibinfo{title}{{The Victoria-Regina Stellar Models: Evolutionary Tracks and Isochrones for a Wide Range in Mass and Metallicity that Allow for Empirically Constrained Amounts of Convective Core Overshooting},} \apjs, 162, 375, \dodoi{10.1086/498451}

\bibitem[{L.~C. {Vargas} {et~al.}(2014){Vargas}, {Gilbert}, {Geha}, {Tollerud}, {Kirby}, \& {Guhathakurta}}]{Vargas2014}
{Vargas}, L.~C., {Gilbert}, K.~M., {Geha}, M., {et~al.} 2014, \bibinfo{title}{{[{\ensuremath{\alpha}}/Fe] Abundances of Four Outer M31 Halo Stars},} \apjl, 797, L2, \dodoi{10.1088/2041-8205/797/1/L2}

\bibitem[{B.~F. {Williams}(2003){Williams}}]{Williams2003}
{Williams}, B.~F. 2003, \bibinfo{title}{{The Recent Star Formation History of the M31 Disk},} \aj, 126, 1312, \dodoi{10.1086/377347}

\bibitem[{J. {Wojno} {et~al.}(2020){Wojno}, {Gilbert}, {Kirby}, {Escala}, {Beaton}, {Tollerud}, {Majewski}, \& {Guhathakurta}}]{Wojno2020}
{Wojno}, J., {Gilbert}, K.~M., {Kirby}, E.~N., {et~al.} 2020, \bibinfo{title}{{Elemental Abundances in M31: [Fe/H] and [{\ensuremath{\alpha}}/Fe] in M31 Dwarf Galaxies Using Coadded Spectra},} \apj, 895, 78, \dodoi{10.3847/1538-4357/ab8ccb}

\bibitem[{J.~L. {Wojno} {et~al.}(2023){Wojno}, {Gilbert}, {Kirby}, {Escala}, {Guhathakurta}, {Beaton}, {Kalirai}, {Chiba}, \& {Majewski}}]{Wojno2023}
{Wojno}, J.~L., {Gilbert}, K.~M., {Kirby}, E.~N., {et~al.} 2023, \bibinfo{title}{{Elemental Abundances in M31: Individual and Coadded Spectroscopic [Fe/H] and [{\ensuremath{\alpha}}/Fe] throughout the M31 Halo with SPLASH},} \apj, 951, 12, \dodoi{10.3847/1538-4357/acd5d3}

\bibitem[{L. {Yang} {et~al.}(2013){Yang}, {Kirby}, {Guhathakurta}, {Peng}, \& {Cheng}}]{Yang2013}
{Yang}, L., {Kirby}, E.~N., {Guhathakurta}, P., {Peng}, E.~W., \& {Cheng}, L. 2013, \bibinfo{title}{{Measuring Detailed Chemical Abundances from Co-added Medium-resolution Spectra. I. Tests Using Milky Way Dwarf Spheroidal Galaxies and Globular Clusters},} \apj, 768, 4, \dodoi{10.1088/0004-637X/768/1/4}

\end{thebibliography}
\bibliographystyle{aasjournalv7_update}




\end{document}